\pdfoutput=1
\documentclass{article}
\usepackage{arxiv}
\usepackage[T1]{fontenc}
\usepackage[utf8]{inputenc}
\usepackage{float}
\usepackage{graphicx}
\usepackage{subcaption}
\usepackage{booktabs}
\usepackage{tabularx}
\usepackage{adjustbox}
\usepackage{comment}
\usepackage{enumitem}
\usepackage{placeins}
\newlength{\tablewidth}
\AtBeginDocument{\setlength{\tablewidth}{\textwidth}}
\usepackage{amsmath}
\usepackage{amssymb}
\usepackage[hidelinks]{hyperref}


\title{ Mixed-dimensional Matrix and Fiber Fracture in Soft Fibrous Materials}
\renewcommand{\shorttitle}{Mixed-dimensional Matrix and Fiber Fracture in Soft Fibrous Materials}

\author{
L. River Spencer \\
Department of Aerospace Engineering \& Engineering Mechanics,\\
The University of Texas at Austin,\\
Austin, TX 78712 \\
\And
Matthew J. Lohr \\ 
Department of Biomedical Engineering, \\
The University of Texas at Austin,\\
Austin, TX 78712 \\
\And
Chad Landis \\
Department of Aerospace Engineering \& Engineering Mechanics,\\
The Oden Institute for Computational Science and Engineering,\\
The University of Texas at Austin,\\
Austin, TX 78712 \\
\And
Manuel K. Rausch \\
Department of Aerospace Engineering \& Engineering Mechanics,\\
Department of Biomedical Engineering, \\
The Oden Institute for Computational Science and Engineering,\\
The University of Texas at Austin,\\
Austin, TX 78712 \\
\And
Jan N. Fuhg \\
Department of Aerospace Engineering \& Engineering Mechanics,\\
The Oden Institute for Computational Science and Engineering,\\
The University of Texas at Austin,\\
Austin, TX 78712
}

\date{}

\newcommand{\dd}{\mathrm{d}}

\newcommand{\Grad}{\operatorname{Grad}}

\begin{document}

\maketitle
\setlength{\headheight}{23pt}
\raggedbottom

\begin{abstract}
We develop a mixed-dimensional finite-element formulation for modeling fracture in soft materials containing embedded fiber networks. Individual fibers are represented by lower-dimensional beam elements embedded in a nearly incompressible hyperelastic continuum, preserving complex network architectures without requiring each fiber to be resolved volumetrically. Separate phase fields describe fracture in the surrounding matrix and the embedded fibers, allowing both failure processes to evolve independently while interacting through their coupled mechanical response.
 The matrix is modeled using a stabilized displacement–pressure formulation, while the fibers are represented by finite-rotation beam elements that account for damage driven by axial stretching and bending. Both constituents are coupled through translational and rotational penalty terms, allowing forces and moments to be transferred between independent discretizations without requiring the matrix mesh to conform to the fiber geometry. The mechanical response and fracture evolution of both constituents are solved simultaneously within a monolithic framework that enforces irreversible damage. Applications
to increasingly complex fiber networks illustrate how
matrix cracking and fiber failure interact, redistributing
forces as individual fibers degrade.
\end{abstract}


\section{Introduction}
\label{sec:introduction}
Many soft materials contain networks of slender, load-bearing
structures that contribute to their mechanical response. 
These networks may be deliberately introduced as reinforcement, as in fiber-reinforced elastomers, soft actuators, and hydrogel composites \cite{Illeperuma2014,Polygerinos2015,kakaletsis2023mechanics}. In other materials, fibrous networks form part of the underlying microstructure itself, as in collagen-rich biological tissues and fibrin-based networks \cite{holzapfel2000new,garyfallogiannis2023fracture}.
In both cases, the macroscopic response depends not only on the properties of the surrounding material but also
on the deformation, orientation, and connectivity of individual network members
\cite{Ban2016,kakaletsis2023mechanics,Tojaga2023}. 
Beyond their contribution to the intact mechanical response, fibrous networks also influence how damage accumulates and fracture propagates.
Experimental and computational studies of stochastic fiber networks and collagenous tissues have demonstrated that fiber orientation, connectivity, and progressive rupture can govern damage localization, crack propagation, and macroscopic fracture resistance \cite{hadi2012multiscale,vanderheiden2015crack,bircher2019tear,picu2023toughness}.
 This is particularly relevant in engineered systems subjected to repeated deformation, where fatigue-driven crack growth can compromise the durability of otherwise tough hydrogels \cite{Bai2017Fatigue}. Fiber-level mechanisms are also central to failure in biological materials: collagen fiber reorientation and sliding contribute substantially to the tear resistance of soft tissues \cite{Yang2015Skin}, while fracture of fibrin networks influences the mechanical integrity of blood clots under cyclic and rate-dependent loading \cite{Liu2024Fibrin,Kliuchnikov2025Fibrin}. 
Crucially, fracture in soft fibrous materials depends on the mechanical interactions between the surrounding matrix and the embedded fiber network \cite{picu2011mechanics,kakaletsis2023mechanics}. Damage in either constituent redistributes forces through the remaining structure, with these interactions depending on fiber orientation, network connectivity, and constituent stiffness contrast \cite{kakaletsis2023mechanics,Tojaga2023,zhang2013cross,lohr7643646impact}. Multiscale studies have further demonstrated how failure of individual collagen fibers influences the macroscopic failure response of fibrous materials \cite{hadi2012multiscale}. These findings highlight the importance of resolving individual fiber failure and its mechanical interaction with the surrounding matrix.
\begin{figure}[!htbp]
  \centering
  \includegraphics[width=0.85\textwidth]{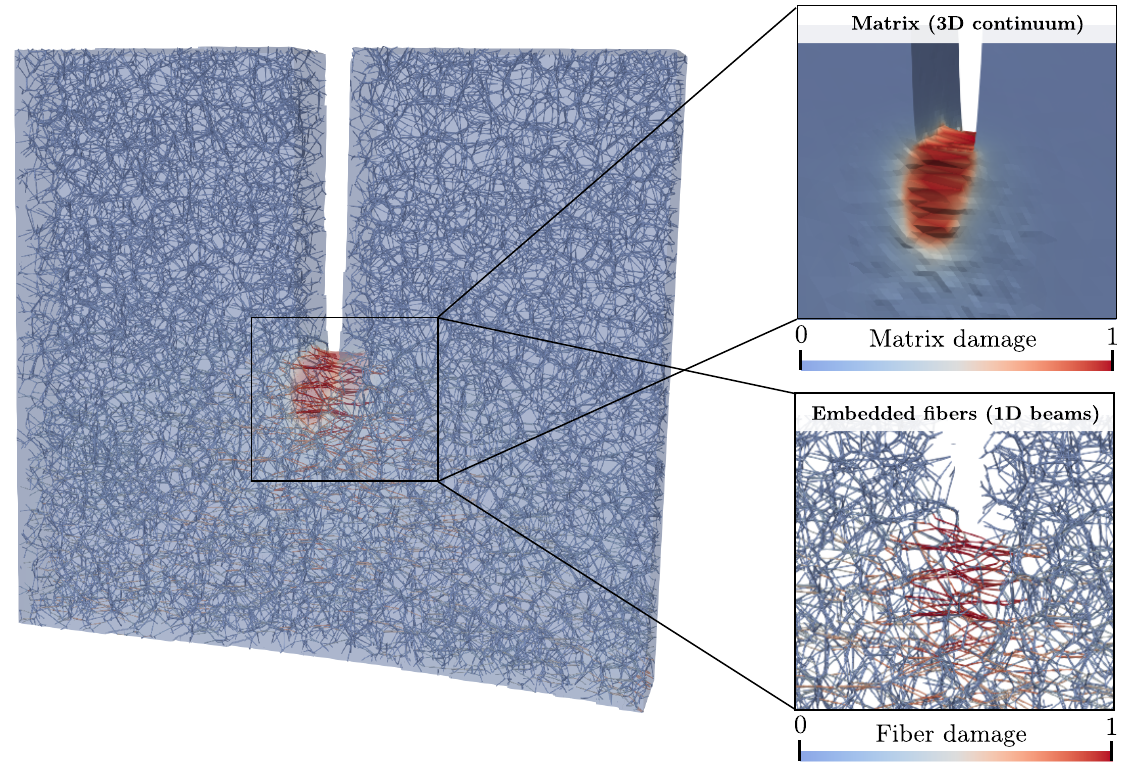}
  \caption{Illustration of coupled matrix and fiber
  fracture in a densely reinforced solid. The matrix
  is represented by a three-dimensional continuum,
  while the embedded fiber network is discretized
  using one-dimensional beam elements. }\label{fig:framework-overview}
\end{figure}

An additional challenge arises from the large deformations characteristic of soft solids. Foundational studies of rubber fracture established energy-based descriptions of crack growth \cite{RivlinThomas1953,LakeThomas1967}, while subsequent investigations demonstrated the importance of nonlinear crack-tip deformation and crack blunting \cite{HuiJagota2003,LongHuiGongBouchbinder2021}. These findings highlight the need to account for finite deformation when modeling fracture in soft materials containing embedded fiber networks.\\ To describe the evolution of fracture, phase-field methods represent cracks through regularized damage fields without explicitly tracking their surfaces. Building on variational descriptions of fracture \cite{FrancfortMarigo1998}, these formulations allow cracks to initiate, propagate, branch, and coalesce without prescribing their paths \cite{Bourdin2000,Borden2012}. Subsequent developments have addressed important aspects of these formulations, including tension--compression energy decompositions, damage irreversibility, and nonlinear solution strategies \cite{Miehe2010,Ambati2015}. Further advances include higher-order regularization to improve the approximation of crack fields \cite{BordenLandis2014}, modified degradation functions that separate the regularization length from the physical process-zone scale in large-structure calculations \cite{LoHughesLandis2023,LoLandis2024Corrigenda}, and extensions to fatigue crack growth \cite{LoBordenRaviChandarLandis2019,LoLandis2024Corrigenda}. Phase-field methods have also been extended to hyperelastic composites \cite{Russ2020}. For materials containing embedded fiber networks, a single damage field cannot describe matrix cracking and fiber rupture independently. Multiple phase fields have therefore been introduced in continuum composite formulations to describe constituent-specific failure mechanisms \cite{Dean2020}. These developments provide a basis for modeling the independent evolution of matrix and fiber damage within a mechanically coupled system. \\
These constituent-specific descriptions must also accommodate the nonlinear and nearly incompressible behavior of soft materials. Finite-deformation phase-field formulations have been developed for hyperelastic solids, including biological tissues \cite{Hesch2014,Gultekin2016}. However, nearly incompressible behavior introduces additional numerical challenges, as intact regions undergo approximately volume-preserving deformation while developing cracks must accommodate substantial local changes in volume. Standard displacement-based finite elements can exhibit volumetric locking under these conditions. Mixed displacement--pressure formulations alleviate this difficulty by introducing an independent pressure field, although equal-order discretizations generally require stabilization. Recent formulations have therefore combined pressure stabilization with damage-dependent relaxation of the volumetric response to model fracture in nearly incompressible hyperelastic solids \cite{Ang2022}. Despite these developments, explicitly representing the geometry and failure of embedded fiber networks presents an additional challenge. Representing individual fibers using three-dimensional solid elements generally requires finely resolved, body-fitted discretizations of the fibers and their interfaces with the surrounding matrix, making complex or densely reinforced networks computationally expensive to model \cite{steinbrecher2020mortar,kakaletsis2023mechanics}. Changes in network architecture may further require reconstruction of the surrounding matrix mesh.\\
These limitations motivate mixed-dimensional finite-element methods, in which slender network members are represented by one-dimensional structural elements embedded within a surrounding continuum. Early formulations established techniques for coupling beams and solids without conforming meshes \cite{mccune2000mixed,sadek2004embedded}. Subsequent mortar and variational approaches have enabled independent matrix and beam discretizations while addressing translational and rotational compatibility and the associated transfer of forces and moments between constituents \cite{steinbrecher2020mortar,khristenko2021multidimensional,Steinbrecher2022}. Such methods have been applied to embedded fiber networks in hyperelastic solids \cite{kakaletsis2023mechanics} and electroactive beam networks in soft dielectric composites \cite{spencer2026mixed}. Separately, phase-field formulations have been developed to describe fracture in reduced-dimensional beam models \cite{Lai2020}. Together, these developments provide the ingredients for a formulation that resolves individual fiber networks and describes fracture in both the surrounding matrix and embedded fibers.\\
Motivated by these challenges, we develop a monolithic mixed-dimensional finite-element framework for modeling
fracture in nearly incompressible hyperelastic solids containing embedded fiber networks.
We combine a stabilized displacement--pressure formulation for the surrounding matrix with
finite-rotation beam elements that resolve the deformation of individual network members.
To describe failure in both constituents, we introduce independent phase fields for matrix cracking and fiber
degradation, while translational and rotational penalty terms provide mechanical coupling across non-matching
discretizations. We solve the resulting mechanical and fracture equations
simultaneously, allowing damage in either constituent to influence the deformation and loading of the other. Figure~\ref{fig:framework-overview} illustrates
the mixed-dimensional representation and the
distinct matrix and fiber damage fields in a
densely reinforced solid.
Finally, we examine the formulation through four numerical demonstrations of increasing complexity, ranging from cyclic loading of a reinforced bar to fracture of an embedded helix, deformation of a
periodic fiber network, and Mode-I fracture of a densely reinforced solid.
These examples illustrate how the formulation captures interacting damage processes and changes in the
load-bearing network without requiring a conforming volumetric mesh of individual fibers.

\section{Methods}
\label{sec:methods}

The proposed framework models a three-dimensional, finite-strain continuum reinforced by an independently discretized network of one-dimensional beams. The matrix and fibers are described by separate mechanical and fracture fields, which interact through translational and rotational penalty coupling between their non-matching meshes \cite{mccune2000mixed,steinbrecher2020mortar,Steinbrecher2022}. The mechanical response and fracture evolution of both constituents are solved simultaneously in a monolithic nonlinear system. While the framework is formulated in three dimensions, a reduced one-dimensional model is also used to examine matrix–fiber load transfer and damage irreversibility.
\subsection{Problem setting and primary fields}
\label{sec:problem-setting}

Consider a deformable body occupying the reference configuration
$\Omega_0\subset\mathbb{R}^3$, with material coordinates
$\mathbf{X}\in\Omega_0$ and displacement field $\mathbf{u}(\mathbf{X})$.
Following a total Lagrangian description, the deformation gradient,
its Jacobian, and the right Cauchy--Green tensor are defined as
\begin{equation}
    \mathbf{F} = \mathbf{I}+\Grad\mathbf{u},
    \qquad
    J = \det\mathbf{F}>0,
    \qquad
    \mathbf{C} = \mathbf{F}^{T}\mathbf{F},
    \label{eq:matrix-kinematics}
\end{equation}
where $\Grad$ denotes differentiation with respect to
the reference coordinates.
Near incompressibility is treated using an independent pressure-like
field $p(\mathbf{X})$ \cite{SimoTaylorPister1985,Ang2022}, while
matrix fracture is described by a scalar phase field
$d(\mathbf{X})\in[0,1]$, where $d=0$ denotes intact material
and $d=1$ complete damage \cite{Miehe2010,Hesch2014}.
The embedded fiber network is represented by a one-dimensional
centerline graph $\Gamma_{f,0}\subset\Omega_0$, parameterized
by the reference arc length $s$ and centerline position
$\mathbf{X}_f(s)$ \cite{mccune2000mixed,steinbrecher2020mortar,Steinbrecher2022}.
The graph carries the fiber displacement $\mathbf{u}_f(s)$,
cross-sectional rotation $\boldsymbol{\theta}_f(s)$, and an
independent phase field $d_f(s)$ describing fiber damage. \\
The complete set of primary unknowns is
\begin{equation}
    \mathbf{q}
    =
    \left[
        \mathbf{u},\,
        p,\,
        d,\,
        \mathbf{y}_f,\,
        d_f
    \right]^T,
    \qquad
    \mathbf{y}_f
    =
    \left(
        \mathbf{u}_f,\,
        \boldsymbol{\theta}_f
    \right),
    \label{eq:unknown-blocks}
\end{equation}
where $\mathbf{y}_f$ collects the beam displacement and rotation
fields. Matrix volume and surface contributions are integrated
over $\Omega_0$ and $\partial\Omega_0$, respectively, while
fiber mechanics, fracture, and matrix--fiber coupling are
integrated along $\Gamma_{f,0}$. The independent discretizations
are coupled through the penalty formulation introduced in
Sec.~\ref{sec:matrix-fiber-coupling}.
\subsection{Nearly incompressible hyperelastic matrix}
\label{sec:matrix-mechanics}
Throughout this work, the matrix is modeled as a nearly incompressible neo-Hookean solid using the stabilized mixed displacement--pressure formulation of Ang et al.~\cite{Ang2022}, building on established approaches to finite-strain hyperelasticity and mixed formulations \cite{spencer2026monolithic,BonetWood2008,SimoTaylorPister1985}. The corresponding mixed potential is
\begin{equation}
    \Psi_m(\mathbf{F},p,d)
    =
    g(d)\psi_{\mu}(\mathbf{F})
    -r(d)p(J-1)
    -\frac{p^2}{2K},
    \label{eq:mixed-potential}
\end{equation}
where $K$ is the volumetric coefficient and the neo-Hookean strain-energy density is defined as
\begin{equation}
    \psi_{\mu}(\mathbf{F})
    =
    \frac{\mu}{2}
    \left(
        \mathbf{F}:\mathbf{F}-3-2\ln J
    \right),
    \label{eq:matrix-energy}
\end{equation}
with shear modulus $\mu$. The stiffness degradation
and pressure relaxation functions are defined as
\begin{equation}
    g(d)=k+(1-k)(1-d)^2,
    \qquad
    r(d)=g(d)^{3/2},
    \label{eq:matrix-degradation}
\end{equation}
where $k>0$ denotes the residual stiffness.
The pressure relaxation function ensures that the
additional volumetric resistance decreases more rapidly
than the neo-Hookean contribution as damage develops,
allowing local volume changes associated with crack opening
while maintaining near incompressibility in undamaged regions
\cite{Ang2022}.

The resulting first Piola--Kirchhoff stress and
local pressure equation are
\begin{align}
    \mathbf{P}
    &=
    g(d)\mu\left(\mathbf{F}-\mathbf{F}^{-T}\right)
    -r(d)pJ\mathbf{F}^{-T},
    \label{eq:first-piola}\\
    0
    &=
    -r(d)(J-1)-\frac{p}{K}.
    \label{eq:pressure-equation}
\end{align}
\subsubsection{Equal-order pressure stabilization}

As the incompressible limit is approached, equal-order displacement and pressure interpolations do not generally satisfy the discrete inf--sup condition and can develop spurious pressure modes \cite{SimoTaylorPister1985,Ang2022}. To retain first-order equal interpolation for the matrix displacement and pressure fields, we adopt the finite-deformation residual-based stabilization of Ang et al.~\cite{Ang2022}. For a matrix element $e$ with reference domain $\Omega_{0,e}$, the stabilization scale is
\begin{equation}
  \tau_e
  =
  c_{\mathrm{stab}}\frac{h_e^2}{2\mu},
  \label{eq:stabilization-scale}
\end{equation}
where $h_e$ is the element diameter and $c_{\mathrm{stab}}$ is dimensionless. The $h_e^2/\mu$ dependence provides the appropriate mesh and material scaling and reduces the stabilization contribution under refinement \cite{Ang2022}.\\
The local pressure residual from Eq.~\eqref{eq:pressure-equation} is supplemented by
\begin{equation}
\begin{split}
  R_p^{\mathrm{stab}}(q_p)
  =
  \sum_e\int_{\Omega_{0,e}}\tau_e\big[&
  -rJ\left(\mathbf{C}^{-1}\Grad(rp)\right)\cdot\Grad q_p
  +\left(\mathbf{P}_{\mu}\Grad g\right)\cdot
    \left(r\mathbf{F}^{-T}\Grad q_p\right)
  \big]\,\dd V,
  \label{eq:pressure-stabilization}
\end{split}
\end{equation}
where $q_p$ is a pressure test function, $\mathbf{P}_{\mu}=\mu(\mathbf{F}-\mathbf{F}^{-T})$, and $g=g(d)$ and $r=r(d)$ are understood locally. The first term controls gradients of the relaxed pressure $rp$ in the finite-deformation metric $\mathbf{C}^{-1}$, while the second accounts for spatial gradients of the degradation across diffuse fracture regions \cite{Ang2022}. In the absence of damage, the formulation reduces to a pressure-gradient stabilization that vanishes for spatially constant pressure fields. 
Because the stabilization depends on the deformation, pressure, and matrix phase field, its derivatives with respect to all three fields are retained in the monolithic tangent. The complete discrete pressure equation is
\begin{equation}
  R_p(q_p)
  =
  \int_{\Omega_0}
  \left[
    -r(d)(J-1)-\frac{p}{K}
  \right]q_p\,\dd V
  +R_p^{\mathrm{stab}}(q_p)
  =
  0
  \label{eq:complete-pressure-residual}
\end{equation}
for every discrete pressure test function. Once stabilization is active, the local pressure relation implied by Eq.~\eqref{eq:pressure-equation} is not expected to hold pointwise; convergence is instead assessed from the assembled discrete pressure residual in Eq.~\eqref{eq:complete-pressure-residual} \cite{Ang2022}. The stabilization is a residual contribution to the discrete pressure equation and is therefore not included in the mechanical-energy quantities reported below.

\subsection{Tensile matrix fracture}
\label{sec:matrix-fracture}
Matrix fracture is described by an AT2 phase-field formulation \cite{Borden2012,AmbrosioTortorelli1990,Bourdin2000,Miehe2010,Ambati2015}. To prevent compressive deformation from directly driving fracture, a spectral tension--compression split of the neo-Hookean energy is adopted \cite{Hesch2014,Tang2019,Ang2022}. Let $\lambda_a$ denote the principal stretches obtained from the eigenvalues of the right Cauchy--Green tensor. The tensile isochoric energy is defined as
\begin{equation}
\psi_{\mu}^{+}
=\frac{\mu}{2}
\sum_{\lambda_a>1}
\left(\lambda_a^2-1-\ln\lambda_a^2\right).
\label{eq:tensile-energy}
\end{equation}
Only principal modes in extension contribute to this energy, while compressive principal modes are excluded from the fracture-driving contribution. The spectral split acts on the local principal deformation rather than the macroscopic loading direction, such that an overall compressive loading state may still generate a nonzero tensile contribution through local principal extensions. The pressure-related contribution is treated separately and activated only during local dilation through the indicator $\chi_{J\geq1}$, which is unity for $J\geq1$ and zero otherwise \cite{Hesch2014,Ang2022}.\\
The effective mechanical driving energy for matrix fracture combines the tensile isochoric contribution with the volumetric contribution associated with dilation. Using the adopted degradation functions, the resulting tensile damage driver is defined as
\begin{equation}
\mathcal{D}_m^+
=
\psi_{\mu}^{+}
-\frac{3}{2}\chi_{J\geq1}\sqrt{g(d)}\,p(J-1).
\label{eq:matrix-tensile-driver}
\end{equation}
The first contribution represents the energy associated with tensile principal deformation, while the second incorporates the influence of volumetric expansion through the mixed pressure field. The dilatational indicator excludes the volumetric contribution during local compression, ensuring that compressive volumetric deformation does not directly promote fracture. The effective driver is evaluated from the current deformation, pressure, and damage fields within the coupled nonlinear formulation.\\
The matrix fracture energy is given by
\begin{equation}
\mathcal{G}_{m}[d]
=G_c\int_{\Omega_0}
\left(
\frac{d^2}{2\ell}
+\frac{\ell}{2}|\Grad d|^2
\right)\,\dd V,
\label{eq:matrix-fracture-energy}
\end{equation}
where $G_c$ is the critical energy-release rate and $\ell$ is the regularization length. The local term penalizes the creation of damage, while the gradient term distributes damage over a diffuse region whose characteristic width is governed by $\ell$ \cite{AmbrosioTortorelli1990,Bourdin2000,Ambati2015}. The matrix discretization must therefore adequately resolve the regularization length when crack paths or fracture loads are interpreted quantitatively \cite{Bourdin2000,Ambati2015,BordenLandis2014}.\\
Combining the effective tensile mechanical driver with the variation of Eq.~\eqref{eq:matrix-fracture-energy} gives the matrix damage residual. For a scalar test function $\eta$,
\begin{equation}
\begin{split}
R_d(\eta)=\int_{\Omega_0}\bigg[
&g'(d)\mathcal{D}_m^+\eta
+\frac{G_c}{\ell}d\eta
+G_c\ell\Grad d\cdot\Grad\eta
\bigg]\,\dd V,
\end{split}
\label{eq:matrix-damage-residual}
\end{equation}
where the derivatives of the degradation functions satisfy
\begin{equation}
g'(d)=-2(1-k)(1-d),
\qquad
r'(d)=\frac{3}{2}\sqrt{g(d)}\,g'(d).
\end{equation}
Substitution of Eq.~\eqref{eq:matrix-tensile-driver} recovers the separate isochoric and pressure-dependent contributions to the original damage residual. The mechanical term promotes damage evolution through the selected tensile contributions, while the local and gradient terms provide resistance through the regularized fracture energy. Their competition governs the development and spatial localization of matrix damage, subject to the imposed irreversibility constraint.\\
Although only the selected tensile contributions drive the phase field, damage degrades the full matrix mechanical response. The neo-Hookean contribution scales with $g(d)$, while the additional volumetric resistance scales with $g(d)^3$. Consequently, matrix damage reduces stiffness under both tensile and compressive deformation. Unlike formulations that explicitly account for unilateral crack closure or frictional contact \cite{Amor2009Unilateral,Hu2020Frictionless,FeiChoo2020Contact}, the present formulation does not introduce a separate crack-face contact response. The fracture-driving contribution therefore differs from the derivative of the complete degraded mechanical energy, resulting in a hybrid formulation in which the mechanical and damage residuals are defined through distinct constitutive contributions. The coupled residuals and their consistent linearization are treated directly in the monolithic solution procedure described in Sec.~\ref{sec:nonlinear-solution} \cite{Ambati2015}.\\
Matrix damage is irreversible between accepted load states,
\begin{equation}
d_{n+1}\geq d_n,
\qquad
0\leq d_{n+1}\leq1,
\label{eq:matrix-irreversibility}
\end{equation}
where $d_n$ is the phase field at the preceding accepted equilibrium state. These nodal bounds are enforced together with the monolithic nonlinear system using the active-set procedure described in Sec.~\ref{sec:nonlinear-solution}. On boundaries without prescribed damage or periodic constraints, the matrix phase field satisfies the natural zero-flux condition
\begin{equation}
\Grad d\cdot\mathbf{N}=0,
\end{equation}
where $\mathbf{N}$ is the outward reference normal.\\

\subsection{Finite-rotation fiber beams}
\label{sec:fiber-beam-mechanics}

The embedded reinforcement is modeled as a network of one-dimensional finite-rotation beams that support axial deformation, transverse shear, bending, and torsion while remaining geometrically independent of the surrounding matrix \cite{spencer2026monolithic,kakaletsis2023mechanics,steinbrecher2020mortar}. The beam kinematics follow geometrically exact spatial rod formulations in which the centerline and cross-sectional orientation evolve independently under finite deformation \cite{Simo1985,SimoVuQuoc1988}. For a reference centerline $\mathbf{X}_f(s)$ parameterized by reference arc length $s$, each initially straight beam segment is equipped with an orthonormal reference frame $\mathbf{E}_0=[\mathbf{t}_0,\mathbf{n}_0,\mathbf{b}_0]$. The current centerline is
$
  \mathbf{x}_f(s)
  =
  \mathbf{X}_f(s)+\mathbf{u}_f(s)$, 
where $\mathbf{u}_f(s)$ is the centerline displacement. The cross-sectional orientation is described by the rotation $\mathbf{R}_f(s)\in\mathrm{SO}(3)$, which is parameterized locally by a rotation vector $\boldsymbol{\vartheta}(s)$ through $\mathbf{R}_f=\exp([\boldsymbol{\vartheta}]_{\times})$, with $[\cdot]_{\times}$ denoting the skew-symmetric operator \cite{Simo1985,SimoVuQuoc1988}.\\
For an initially straight beam segment with orthonormal
reference frame
$\mathbf{E}_0=[\mathbf{t}_0,\mathbf{n}_0,\mathbf{b}_0]$,
the generalized strain and curvature are
\begin{equation}
  \boldsymbol{\gamma}
  =
  \mathbf{R}_f^T
  \left(
    \mathbf{e}_t+\mathbf{E}_0^T\mathbf{u}_{f,s}
  \right)
  -\mathbf{e}_t,
  \qquad
  \boldsymbol{\kappa}
  =
  \operatorname{axl}
  \left(
    \mathbf{R}_f^T\mathbf{R}_{f,s}
  \right)
  =
  \mathbf{J}_r(\boldsymbol{\vartheta})
  \boldsymbol{\vartheta}_{,s},
  \label{eq:fiber-generalized-strains}
\end{equation}
Here $(\cdot)_{,s}=\partial(\cdot)/\partial s$ denotes differentiation with respect to reference arc length and $\mathbf{e}_t=(1,0,0)^T$ is the axial direction of the reference beam frame. 
The generalized strain and curvature vectors are written as $\boldsymbol{\gamma}=(\gamma_t,\gamma_n,\gamma_b)^T$ and $\boldsymbol{\kappa}=(\kappa_t,\kappa_n,\kappa_b)^T$, where $\gamma_t$ is the axial strain, $\gamma_n$ and $\gamma_b$ are the transverse shear strains, $\kappa_t$ is the twist curvature, and $\kappa_n$ and $\kappa_b$ are the bending curvatures \cite{Simo1985,SimoVuQuoc1988}. The operator $\operatorname{axl}(\cdot)$ extracts the axial vector of a skew-symmetric tensor, while $\mathbf{J}_r(\boldsymbol{\vartheta})$ is the right Jacobian. 
Each beam segment is assigned an independent reference frame, with $\mathbf{t}_0$ aligned with the straight line connecting its endpoints in the undeformed configuration and the transverse directions completing an orthonormal basis. A curved fiber is therefore represented as a piecewise-straight, stress-free reference polygon, so $\boldsymbol{\kappa}$ measures the change in section orientation relative to each segment's own straight reference state. 
The axial constitutive response is constructed from a compressible neo-Hookean law and calibrated so that the prescribed Young's modulus $E_f$ is recovered in the infinitesimal axial limit. For a circular cross-section of radius $r_f$ and area $A_f=\pi r_f^2$, define
\begin{equation}
  \mu_f
  =
  \frac{E_f}{2(1+\nu_f)},
  \qquad
  \lambda_{\mathrm{L},f}
  =
  \frac{E_f\nu_f}
  {(1+\nu_f)(1-2\nu_f)},
  \qquad
  \alpha_f
  =
  \frac{E_f}
  {\lambda_{\mathrm{L},f}+2\mu_f}.
  \label{eq:fiber-moduli}
\end{equation}
Restricting the finite-strain constitutive response to an axial stretch $\lambda_f$ gives the energy per unit reference centerline length \cite{SimoPister1984}
\begin{equation}
  \psi_f^{\mathrm{ax}}(\lambda_f)
  =
  A_f\alpha_f
  \left[
    \frac{\mu_f}{2}(\lambda_f^2-1)
    -\mu_f\ln\lambda_f
    +\frac{\lambda_{\mathrm{L},f}}{2}(\ln\lambda_f)^2
  \right].
  \label{eq:fiber-energy}
\end{equation}
The scaling factor $\alpha_f$ ensures that the
corresponding axial force
$N_f=\partial\psi_f^{\mathrm{ax}}/\partial\lambda_f$
satisfies
\begin{equation}
  \left.
  \frac{\partial N_f}{\partial\lambda_f}
  \right|_{\lambda_f=1}
  =
  E_fA_f.
  \label{eq:fiber-tangent-calibration}
\end{equation} 
To incorporate bending and torsion, we consider a
material point with cross-sectional coordinates
$(x_n,x_b)$. Its section strains are
\begin{equation}
  a_t
  =
  \gamma_t+\kappa_nx_b-\kappa_bx_n,
  \qquad
  a_n
  =
  \gamma_n-\kappa_tx_b,
  \qquad
  a_b
  =
  \gamma_b+\kappa_tx_n.
  \label{eq:fiber-section-strains}
\end{equation}
The local axial stretch is $1+a_t$, so the axial centerline deformation and bending both contribute to the extension of individual section points. The transverse measures $a_n$ and $a_b$ combine centerline shear with the kinematic contribution of twist. With $w_f(\lambda)=\psi_f^{\mathrm{ax}}(\lambda)/A_f$, the undamaged beam energy per unit reference centerline length is
\begin{equation}
  \psi_f^{\mathrm{beam}}
  =
  \int_{A_f}
  \left[
    w_f(1+a_t)
    +
    \frac{\mu_f}{2}
    \left(
      a_n^2+a_b^2
    \right)
  \right]
  \,\dd A.
  \label{eq:spatial-beam-energy}
\end{equation}
This section-integrated response combines finite axial deformation, transverse shear, bending, and torsion within the same beam description. In the infinitesimal limit, Eq.~\eqref{eq:spatial-beam-energy} recovers the classical axial, shear, bending, and torsional stiffnesses $E_fA_f$, $\mu_fA_f$, $E_fI_f$, and $\mu_fJ_f$, with $I_f=\pi r_f^4/4$ and $J_f=2I_f$ \cite{Simo1985,SimoVuQuoc1988}.\\
The mechanical potential of the embedded fibers is
\begin{equation}
  \Pi_f^{\mathrm{mech}}[\mathbf{y}_f]
  =
  \int_{\Gamma_{f,0}}
  \psi_f^{\mathrm{beam}}
  \,\dd s.
  \label{eq:fiber-mechanical-potential}
\end{equation}
The constitutive response requires positive axial stretch throughout the complete circular section. Because $a_t$ is affine in $(x_n,x_b)$, the minimum section stretch is
\begin{equation}
  \lambda_{\mathrm{sec,min}}
  =
  1+\gamma_t
  -r_f\sqrt{\kappa_n^2+\kappa_b^2} >0\, .
  \label{eq:minimum-section-stretch}
\end{equation}
At rigid fiber network junctions, incident branches share their centerline position and cross-sectional rotation, allowing force and moment to pass between neighboring members. Hinged junctions instead enforce common translation while leaving the branch rotations independent, giving an ideal joint that transfers forces but not moments. The corresponding treatment of the fiber phase field is introduced in Sec.~\ref{sec:fiber-fracture}. The finite-element interpolation of the centerline and rotation fields, and admissibility checks during the nonlinear solve are collected with the remaining discretization choices in Sec.~\ref{sec:discretization}.\\

\subsection{Fiber phase-field fracture on the embedded graph}
\label{sec:fiber-fracture}

Fiber rupture is described by an independent scalar phase field
$d_f:\Gamma_{f,0}\rightarrow[0,1]$. The matrix and fiber phase
fields evolve separately and interact only through their
mechanical response and matrix--fiber load transfer, rather
than through direct fracture coupling \cite{Dean2020}.\\
Following the phase-field treatment of beam fracture
\cite{Lai2020}, fiber damage is driven by the tensile axial
energy across the beam section. Because the section strain
$a_t$ in Eq.~\eqref{eq:fiber-section-strains} includes both
centerline extension and bending, tensile deformation on
one side of a bent section can initiate damage even without
centerline extension. To exclude compressed section points
while maintaining smoothness, we introduce the regularized
tensile indicator
\begin{equation}
  \mathcal{H}_{\varepsilon_s}(a_t)=
  \begin{cases}
    0, & a_t\leq0,\\
    6\xi^5-15\xi^4+10\xi^3,
      & 0<a_t<\varepsilon_s,\\
    1, & a_t\geq\varepsilon_s,
  \end{cases}
  \qquad
  \xi=\frac{a_t}{\varepsilon_s},
  \qquad
  \varepsilon_s=10^{-6},
  \label{eq:fiber-tensile-switch}
\end{equation}
where $\varepsilon_s$ controls the width of the $C^2$
transition around zero strain. 
The resulting tensile fracture-driving energy per unit reference centerline length is
\begin{equation}
  \psi_f^+
  =
  \int_{A_f}
  \mathcal{H}_{\varepsilon_s}(a_t)
  w_f(1+a_t)\,\dd A,
  \label{eq:fiber-energy-split}
\end{equation}
where $w_f(\lambda)=\psi_f^{\mathrm{ax}}(\lambda)/A_f$. Only section points in axial extension contribute to $\psi_f^+$. Bending therefore contributes through the tensile section strain, whereas transverse shear, torsion, and axial compression do not directly drive damage. 
Fiber damage degrades the complete beam response through
\begin{equation}
  g_f(d_f)
  =
  k_f+(1-k_f)(1-d_f)^2,
  \label{eq:fiber-degradation}
\end{equation}
where $k_f>0$ is a residual stiffness parameter. Since $d_f$
is defined along the centerline, it describes damage
of the entire beam section, reducing axial, shear,
bending, and torsional stiffnesses together.
The formulation therefore represents whole-section
rupture rather than progressive fracture across the
fiber diameter \cite{Lai2020}.\\
Fracture along the embedded fibers is regularized
through a one-dimensional AT2 functional
\cite{Borden2012,Bourdin2000,Ambati2015},
\begin{equation}
  \Pi_f^{\mathrm{frac}}
  =
  \int_{\Gamma_{f,0}}
  G_{c,f}
  \left[
    \frac{d_f^2}{2\ell_f}
    +
    \frac{\ell_f}{2}
    \left(
      \frac{\partial d_f}{\partial s}
    \right)^2
  \right]
  \,\dd s,
  \label{eq:fiber-fracture}
\end{equation}
where $G_{c,f}$ is the fiber fracture-energy parameter and $\ell_f$ controls the regularization length along the reference centerline. 
Combining the tensile fracture driver with the
AT2 regularization gives the fiber damage residual
\begin{equation}
  R_{d_f}(\eta_f)
  =
  \int_{\Gamma_{f,0}}
  \left[
    g_f'(d_f)\psi_f^+\eta_f
    +
    \frac{G_{c,f}}{\ell_f}d_f\eta_f
    +
    G_{c,f}\ell_f
    \frac{\partial d_f}{\partial s}
    \frac{\partial\eta_f}{\partial s}
  \right]
  \,\dd s,
  \label{eq:fiber-damage-residual}
\end{equation}
where $g_f'(d_f)=-2(1-k_f)(1-d_f)$. 
As in the matrix formulation, the tensile energy
drives fracture while damage degrades the complete
mechanical response.  \\
The fiber network topology determines the natural conditions associated with the gradient term. At a rigid junction, $d_f$ is continuous across the incident branches and the outward fracture fluxes satisfy
\begin{equation}
  \sum_{e\in\mathcal{E}(v)}
  G_{c,f}\ell_f
  \left.
  \frac{\partial d_f}{\partial n_e}
  \right|_v
  =
  0,
  \label{eq:fiber-junction-flux}
\end{equation}
where $\mathcal{E}(v)$ denotes the branches incident to junction $v$ and $\partial/\partial n_e$ is directed outward along each branch. This Kirchhoff-type condition allows the regularized phase field to extend through a rigid fiber network junction while balancing the gradient fluxes contributed by the connected branches. Hinged junctions instead retain independent branch-end phase fields, each with its own natural zero-flux condition when unconstrained. Free external fiber endpoints likewise satisfy zero flux.\\
Fiber fracture is irreversible between accepted load states,
\begin{equation}
  d_{f,n+1}\geq d_{f,n},
  \qquad
  0\leq d_{f,n+1}\leq1,
  \label{eq:fiber-irreversibility}
\end{equation}
where $d_{f,n}$ is the previously accepted field. These bounds are enforced through the monolithic
active-set procedure in Sec.~\ref{sec:nonlinear-solution}.

\subsection{Non-matching line-to-volume penalty coupling}
\label{sec:matrix-fiber-coupling}

The matrix continuum and embedded fibers are discretized independently and interact through a line-to-volume penalty coupling along the reference fiber centerline. This construction transfers force and moment between non-matching three- and one-dimensional fields without requiring the matrix mesh to conform to the fiber geometry \cite{spencer2026monolithic,mccune2000mixed,sadek2004embedded,steinbrecher2020mortar,khristenko2021multidimensional,kakaletsis2023mechanics}. The reinforcement is superposed on the host continuum: matrix material remains present throughout $\Omega_0$, including the nominal beam volume, and the beam contributes an additional reduced-dimensional mechanical response. A useful measure of the resulting reinforcement density is the nominal superposed fiber fraction
\begin{equation}
  \phi_f^{\mathrm{sup}}
  =
  \frac{A_fL_{\mathrm{tot}}}{V_0}
  =
  \frac{\pi r_f^2L_{\mathrm{tot}}}{V_0},
  \label{eq:superposed-fiber-fraction}
\end{equation}
where $L_{\mathrm{tot}}$ is the total reference centerline length, and $V_0$ is the reference matrix volume. Because no matrix volume is removed around the beams, $\phi_f^{\mathrm{sup}}$ characterizes the amount of added reinforcement rather than a partition of the body into disjoint matrix and fiber phases.\\
Translational compatibility is imposed along the embedded fibers through the relative displacement
\begin{equation}
  \mathbf{g}_{\mathrm{int}}(s)
  =
  \mathbf{u}_f(s)
  -
  \mathbf{u}(\mathbf{X}_f(s)),
  \label{eq:interface-gap}
\end{equation}
where both fields are evaluated at the reference fiber position $\mathbf{X}_f(s)$. 
Let $\mathbf{t}_0$ denote the reference fiber tangent. The translational penalty stiffness may distinguish motion parallel to and transverse to the fiber through
\begin{equation}
  \mathbf{K}_{\mathrm{int}}
  =
  k_t\,\mathbf{t}_0\otimes\mathbf{t}_0
  +
  k_n
  \left(
    \mathbf{I}-\mathbf{t}_0\otimes\mathbf{t}_0
  \right),
  \label{eq:interface-stiffness-tensor}
\end{equation}
where $k_t$ and $k_n$ are the axial and transverse penalty stiffnesses. The calculations considered here use $k_t=k_n=k_{\mathrm{int}}$, giving the isotropic interaction force per unit reference length
\begin{equation}
  \mathbf{t}_{\mathrm{int}}
  =
  \mathbf{K}_{\mathrm{int}}\mathbf{g}_{\mathrm{int}}
  =
  k_{\mathrm{int}}\mathbf{g}_{\mathrm{int}}.
  \label{eq:interface-traction}
\end{equation}
The corresponding penalty potential is
\begin{equation}
  \Pi_{\mathrm{int}}
  =
  \frac{1}{2}
  \int_{\Gamma_{f,0}}
  k_{\mathrm{int}}
  \|\mathbf{g}_{\mathrm{int}}\|^2
  \,\dd s,
  \label{eq:interface-energy}
\end{equation}
where $k_{\mathrm{int}}$ is the translational penalty
stiffness \cite{mccune2000mixed,sadek2004embedded,
steinbrecher2020mortar,khristenko2021multidimensional}.
The resulting interaction force per unit reference length,
$\mathbf{t}_{\mathrm{int}}
=k_{\mathrm{int}}\mathbf{g}_{\mathrm{int}}$,
acts equally and oppositely on the matrix and fibers.
Following Steinbrecher et al.~\cite{Steinbrecher2022},
we additionally couple the beam directors to the local
matrix rotation $\mathbf{R}_m$, obtained from the
polar decomposition
$\mathbf{F}=\mathbf{R}_m\mathbf{U}_m$.
The matrix target directors are
$\widehat{\mathbf{n}}_m=\mathbf{R}_m\mathbf{n}_0$
and
$\widehat{\mathbf{b}}_m=\mathbf{R}_m\mathbf{b}_0$,
while the corresponding beam directors are
$\mathbf{d}_n=\mathbf{R}_f\mathbf{n}_0$
and
$\mathbf{d}_b=\mathbf{R}_f\mathbf{b}_0$.\\
Rotational compatibility is enforced through
\begin{equation}
  \Pi_{\mathrm{rot}}
  =
  \frac{1}{4}
  \int_{\Gamma_{f,0}}
  k_{\theta}
  \left(
    \|\mathbf{d}_n-\widehat{\mathbf{n}}_m\|^2
    +
    \|\mathbf{d}_b-\widehat{\mathbf{b}}_m\|^2
  \right)
  \,\dd s,
  \label{eq:rotational-interface-energy}
\end{equation}
where $k_{\theta}$ is the rotational penalty stiffness.
Using both transverse directors ensures that the
coupling does not depend on an arbitrarily selected
transverse direction for the circular cross-section
\cite{Steinbrecher2022}.
The penalty vanishes under a compatible common
rigid rotation and reduces to a quadratic rotational
penalty for small relative rotations.

Both penalty contributions approximate a perfectly
bonded matrix--fiber interface and remain independent
of the constituent phase fields.
Matrix and fiber damage therefore affect load transfer
through their evolving mechanical responses rather
than through degradation of the coupling itself.
The finite-element transfer, penalty scaling, and
quadrature are detailed in Sec.~\ref{sec:discretization}.

\subsection{Residual construction, linearization, and finite-element discretization}
\label{sec:discretization}

The coupled formulation combines energy-derived mechanical
contributions with the stabilized pressure equation and
the tensile fracture residuals introduced above.
The matrix and degraded fiber mechanical potentials are
\begin{equation}
  \Pi_m^{\mathrm{mech}}
  =
  \int_{\Omega_0}
  \Psi_m(\mathbf F,p,d)\,\dd V,
  \qquad
  \Pi_{f,\mathrm{deg}}^{\mathrm{mech}}
  =
  \int_{\Gamma_{f,0}}
  g_f(d_f)\psi_f^{\mathrm{beam}}(\mathbf y_f)\,\dd s.
  \label{eq:constituent-mechanical-energies}
\end{equation}
The mechanical equilibrium equations follow from these
potentials together with the translational and rotational
coupling terms. The pressure and fracture equations additionally
contain the stabilization and tensile driving-force contributions,
respectively.\\
The five coupled residuals are therefore
\begin{align}
  \mathcal G_u(\delta\mathbf u)
  &=
  D_{\mathbf u}
  \left[
    \Pi_m^{\mathrm{mech}}
    +\Pi_{\mathrm{int}}
    +\Pi_{\mathrm{rot}}
    -\mathcal W_{\mathrm{ext}}
  \right][\delta\mathbf u],
  \label{eq:matrix-mechanical-variation}\\
  \mathcal G_p(q_p)
  &=
  D_p\Pi_m^{\mathrm{mech}}[q_p]
  +R_p^{\mathrm{stab}}(q_p),
  \label{eq:pressure-variation}\\
  \mathcal G_{y_f}(\delta\mathbf y_f)
  &=
  D_{\mathbf y_f}
  \left[
    \Pi_{f,\mathrm{deg}}^{\mathrm{mech}}
    +\Pi_{\mathrm{int}}
    +\Pi_{\mathrm{rot}}
    -\mathcal W_{\mathrm{ext}}
  \right][\delta\mathbf y_f],
  \label{eq:fiber-mechanical-variation}\\
  \mathcal G_d(\eta)
  &=R_d(\eta),
  \qquad
  \mathcal G_{d_f}(\eta_f)=R_{d_f}(\eta_f).
  \label{eq:fracture-residual-definitions}
\end{align}
Here $D_{\mathbf z}$ denotes the directional derivative
with respect to the field $\mathbf z$. Because the mechanical response is degraded by the full
constituent energies, whereas fracture is driven only by
their selected tensile contributions, the resulting hybrid
system cannot be derived from a single global potential
\cite{Ambati2015}.\\
After assembly, the discrete unknown and residual vectors are written as
\begin{equation}
  \mathbf q
  =
  \begin{bmatrix}
    \mathbf u & p & d & \mathbf y_f & d_f
  \end{bmatrix}^{T},
  \qquad
  \mathbf R(\mathbf q)
  =
  \begin{bmatrix}
    \mathbf R_u &
    \mathbf R_p &
    \mathbf R_d &
    \mathbf R_{y_f} &
    \mathbf R_{d_f}
  \end{bmatrix}^{T},
  \label{eq:block-residual}
\end{equation}
where the five blocks represent matrix equilibrium, the stabilized pressure equation, matrix fracture, beam equilibrium including matrix--fiber coupling, and fiber fracture. The monolithic tangent is
\begin{equation}
  \mathbf K(\mathbf q)
  =
  \frac{\partial\mathbf R}{\partial\mathbf q}
  =
  \begin{bmatrix}
    \mathbf K_{uu} & \mathbf K_{up} & \mathbf K_{ud} & \mathbf K_{uy_f} & \mathbf K_{ud_f}\\
    \mathbf K_{pu} & \mathbf K_{pp} & \mathbf K_{pd} & \mathbf K_{py_f} & \mathbf K_{pd_f}\\
    \mathbf K_{du} & \mathbf K_{dp} & \mathbf K_{dd} & \mathbf K_{dy_f} & \mathbf K_{dd_f}\\
    \mathbf K_{y_fu} & \mathbf K_{y_fp} & \mathbf K_{y_fd} & \mathbf K_{y_fy_f} & \mathbf K_{y_fd_f}\\
    \mathbf K_{d_fu} & \mathbf K_{d_fp} & \mathbf K_{d_fd} & \mathbf K_{d_fy_f} & \mathbf K_{d_fd_f}
  \end{bmatrix},
  \label{eq:monolithic-tangent}
\end{equation}
with $\mathbf K_{AB}=\partial\mathbf R_A/\partial\mathbf q_B$. 
All nonzero cross derivatives generated by the mixed matrix law, constituent degradation, beam mechanics, fracture fields, and matrix--fiber coupling are retained. Linearization about the current state gives
\begin{equation}
  \mathbf R
  \left(
    \mathbf q^{(i)}+\Delta\mathbf q^{(i)}
  \right)
  \approx
  \mathbf R(\mathbf q^{(i)})
  +
  \mathbf K(\mathbf q^{(i)})
  \Delta\mathbf q^{(i)}.
  \label{eq:residual-linearization}
\end{equation}
The resulting nonlinear system is solved using the
monolithic procedure described in
Sec.~\ref{sec:nonlinear-solution}.
The matrix and fibers are discretized independently, preserving the nonmatching character of the mixed-dimensional formulation \cite{spencer2026monolithic,steinbrecher2020mortar,khristenko2021multidimensional}. Matrix displacement, pressure, and phase field use equal-order first-order interpolation with the pressure stabilization introduced in Sec.~\ref{sec:matrix-mechanics}. The fibers' displacement, rotation, and phase field are interpolated independently along each beam element. With matrix and fiber shape functions $N_a^m$ and $N_a^f$,
\begin{equation}
  \begin{aligned}
    \mathbf u^h&=\sum_aN_a^m\mathbf u_a,
    & p^h&=\sum_aN_a^m p_a,
    & d^h&=\sum_aN_a^m d_a,\\
    \mathbf u_f^h&=\sum_aN_a^f\mathbf u_{f,a},
    & \boldsymbol\theta_f^h&=\sum_aN_a^f\boldsymbol\theta_{f,a},
    & d_f^h&=\sum_aN_a^f d_{f,a}.
  \end{aligned}
  \label{eq:discrete-fields}
\end{equation}
For spatial beams, the interpolated global rotation
vector is expressed in the local element frame as
$\boldsymbol\vartheta^h
=\mathbf E_0^T\boldsymbol\theta_f^h$,
with $\mathbf R_f^h=\exp([\boldsymbol\vartheta^h]_\times)$. 
Matrix and fiber contributions are integrated on their
respective meshes. For the nonmatching coupling, each
fiber integration point evaluates the matrix interpolation
in the host element containing its reference position
$\mathbf X_f(s)$, transferring the coupling contributions
between the corresponding degrees of freedom
\cite{steinbrecher2020mortar,khristenko2021multidimensional}.
For the line-to-volume coupling, we use two-point
Gauss quadrature on each beam element without
subdividing the integration at matrix-element boundaries.\\
The finite translational and rotational penalties are scaled relative to the local beam stiffness. For a fiber element of reference length $L_e$, define
\begin{equation}
  L_c=\max(L_e,L_{\min}),
  \qquad
  k_{\mathrm{int}}
  =
  c_b\frac{E_fA_f}{L_c^2},
  \qquad
  k_{\theta}
  =
  c_{\theta}\frac{\mu_fJ_f}{L_c^2},
  \label{eq:discrete-penalty-stiffness}
\end{equation}
where $c_b$ and $c_\theta$ are dimensionless penalty factors. The length floor $L_{\min}$ prevents the penalty stiffness from increasing without bound on short fibers. For $L_e\geq L_{\min}$, integration over the element gives coupling stiffnesses that scale with the corresponding beam stiffnesses, while larger penalty factors reduce compatibility error at the expense of conditioning \cite{steinbrecher2020mortar,khristenko2021multidimensional}.\\
At a fiber integration point, let $\mathbf N_m$ and $\mathbf N_f$ collect the matrix and fiber displacement interpolation matrices. The discrete displacement gap is
\begin{equation}
  \mathbf g_{\mathrm{int},h}
  =
  \mathbf N_f\mathbf u_f^e
  -
  \mathbf N_m\mathbf u_m^e,
  \label{eq:discrete-interface-gap}
\end{equation}
and a contribution with integration weight $w_q$ gives
\begin{equation}
  \mathbf R_{\mathrm{int}}^q
  =
  w_q
  \begin{bmatrix}
    -\mathbf N_m^T\mathbf t_{\mathrm{int}}\\
     \mathbf N_f^T\mathbf t_{\mathrm{int}}
  \end{bmatrix},
  \qquad
  \mathbf K_{\mathrm{int}}^q
  =
  w_q
  \begin{bmatrix}
    \mathbf N_m^T\mathbf K_{\mathrm{int}}\mathbf N_m &
    -\mathbf N_m^T\mathbf K_{\mathrm{int}}\mathbf N_f\\
    -\mathbf N_f^T\mathbf K_{\mathrm{int}}\mathbf N_m &
    \mathbf N_f^T\mathbf K_{\mathrm{int}}\mathbf N_f
  \end{bmatrix}.
  \label{eq:discrete-interface-system}
\end{equation}
The equal-and-opposite residual contributions generate the matrix--matrix, matrix--fiber, fiber--matrix, and fiber--fiber tangent blocks. The rotational coupling is assembled analogously from the director mismatch in Eq.~\eqref{eq:rotational-interface-energy}, retaining its dependence on both the matrix and beam rotations.\\
The tensile matrix energy $\psi_\mu^+$ is evaluated from the
principal stretches of $\mathbf C$. Its derivatives are computed
numerically to avoid explicitly differentiating the eigenvectors,
which are non-unique at repeated eigenvalues. The spatial beam
and matrix--fiber coupling contributions are consistently
linearized, including the derivatives of the regularized tensile
switch in Eq.~\eqref{eq:fiber-tensile-switch}.

\subsection{Monolithic nonlinear solution}
\label{sec:nonlinear-solution}

Matrix and fiber irreversibility are enforced relative to the previously accepted phase fields. Let $\mathcal V_d^h$ and $\mathcal V_{d_f}^h$ denote the discrete matrix and fiber phase-field spaces after essential or periodic constraints. At load step $n+1$, the admissible sets are
\begin{equation}
\begin{aligned}
  \mathcal K_d^n
  &=
  \left\{
    \widehat d\in\mathcal V_d^h:
    d_n\leq\widehat d\leq1
  \right\},
  \\
  \mathcal K_{d_f}^n
  &=
  \left\{
    \widehat d_f\in\mathcal V_{d_f}^h:
    d_{f,n}\leq\widehat d_f\leq1
  \right\},
  \label{eq:damage-admissible-sets}
\end{aligned}
\end{equation}
where $d_n$ and $d_{f,n}$ are the previously accepted
phase fields.
The corresponding variational inequalities are
\begin{align}
  \mathcal G_d(\widehat d-d)
  &\geq0, \quad
  \forall\widehat d\in\mathcal K_d^n,
  \label{eq:matrix-damage-inequality}\\
  \mathcal G_{d_f}(\widehat d_f-d_f)
  &\geq0, \quad
  \forall\widehat d_f\in\mathcal K_{d_f}^n.
  \label{eq:fiber-damage-inequality}
\end{align}
These constraints are enforced using a primal--dual active-set
procedure within the monolithic Newton system
\cite{HeisterWheelerWick2015,FarrellMaurini2017}.
Degrees of freedom at active damage bounds are held fixed,
while the remaining mechanical, pressure, and fracture
unknowns are solved simultaneously. The active set is
updated during the nonlinear iterations.\\
At Newton iteration $i$, Eq.~\eqref{eq:residual-linearization} gives
\begin{equation}
  \mathbf K(\mathbf q^{(i)})
  \Delta\mathbf q^{(i)}
  =
  -\mathbf R(\mathbf q^{(i)}),
  \qquad
  \mathbf q^{(i+1)}
  =
  \mathbf q^{(i)}
  +
  \alpha\Delta\mathbf q^{(i)},
  \label{eq:newton}
\end{equation}
where $\alpha$ is selected by residual-based backtracking. Trial states outside the constitutive domains of the matrix or beam are rejected, including non-positive matrix Jacobians and non-positive beam section stretches. After rebuilding the active set and projected residual, convergence requires
\begin{equation}
  \|\mathbf R^{(i)}\|_2
  \leq
  \max\left(
    \varepsilon_{\mathrm{abs}},
    \varepsilon_{\mathrm{rel}}
    \|\mathbf R^{(0)}\|_2
  \right).
  \label{eq:newton-convergence}
\end{equation}
The same projected residual is used to assess the line search. \\
The fracture contributions are evaluated from the current
mechanical state throughout the nonlinear iterations, without
introducing frozen history fields. Matrix and fiber damage
histories are updated only after the complete coupled system
has converged \cite{FarrellMaurini2017}.

\section{Numerical demonstrations}
\label{sec:examples}
We examine the proposed formulation through four numerical
demonstrations, progressing from a reduced verification problem
to increasingly complex three-dimensional fiber networks:

\begin{itemize}[leftmargin=5mm]
    \item \textbf{Cyclic loading of a reinforced bar:}
    We consider a one-dimensional matrix bar containing an
    embedded fiber and subject it to repeated loading and
    unloading. This allows us to examine load transfer between
    the constituents, near-incompressible behavior, and damage
    irreversibility.

    \item \textbf{Extension of an embedded helix:}
    We stretch a helical fiber embedded in a soft matrix and
    vary the fracture resistance of each constituent.
    We examine how these variations affect the sequence of
    matrix and fiber damage and the structural response
    beyond the maximum load.

    \item \textbf{Deformation of a periodic fiber network:}
    We subject a periodic cell containing an irregular
    three-dimensional fiber network to axial extension.
    We investigate how the network deforms through
    stretching, bending, and twisting, and how damage
    develops in the matrix and fibers.

    \item \textbf{Fracture of a densely reinforced solid:}
    We apply Mode-I opening to a notched solid reinforced
    by a large Voronoi fiber network.
    We examine how matrix cracking and fiber damage
    develop together and how local fiber failure changes
    the load-bearing network.
\end{itemize}

For these demonstrations, we select material parameters
representative of soft, nearly incompressible polymer matrices
reinforced by substantially stiffer fibers.
The matrix shear modulus of approximately $\mu=1$ MPa corresponds
to an initial Young's modulus near $3$ MPa, consistent with
reported properties of silicone elastomers such as
polydimethylsiloxane (PDMS) \cite{Li2023PDMS}.
The matrix fracture energy of $G_c=0.2$ N/mm used in the
large-network example is comparable to experimentally measured
cohesive fracture energies of silicone rubber \cite{LiuGent1992}.
For the reinforcement, we use a Young's modulus of
$E_f=100$ MPa, representative of relatively compliant polymer
fibers \cite{Redondo2019}.
While these parameters provide a physically motivated stiffness
contrast, we vary the fracture parameters and regularization
lengths between examples to investigate different damage
mechanisms.

Except for the periodic-cell diagnostics defined in
Section~\ref{sec:periodic-rve-results}, we compute global reactions from
the assembled coupled mechanical residual before essential
boundary constraints are eliminated. These reactions include
the matrix, fiber, and coupling contributions and are conjugate
to the prescribed boundary motion.

\subsection{Cyclic loading of a one-dimensional reinforced bar}
\label{sec:reduced-mixed-cyclic-results}
\begin{figure}[!htbp]
  \centering
  \includegraphics[width=\linewidth]{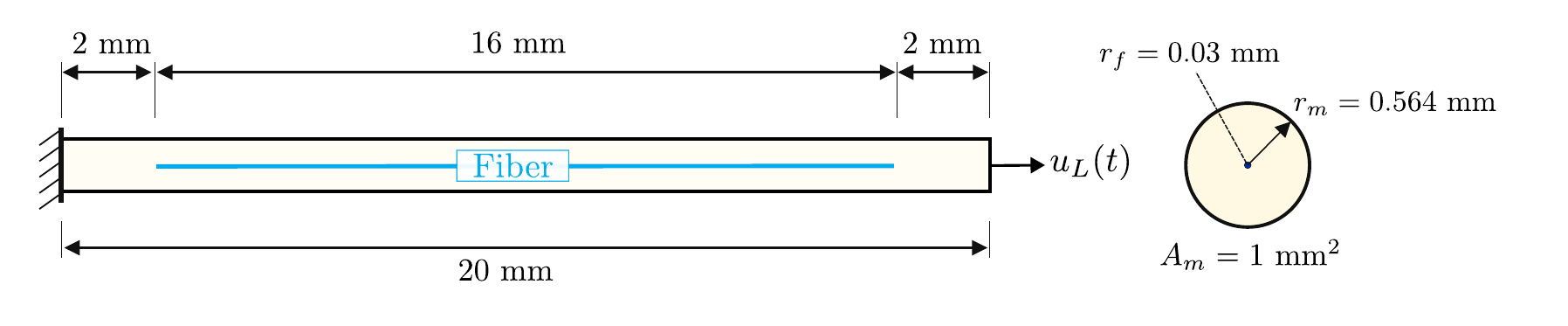}
  \caption{Geometry and cyclic axial loading of the one-dimensional reinforced bar. The enlarged cross-section uses the equivalent matrix radius $r_m=\sqrt{A_m/\pi}$; both fiber ends remain free.}
  \label{fig:reduced-bar-schematic}
\end{figure}

A one-dimensional reinforced bar subjected to cyclic axial loading is considered to examine matrix--fiber load transfer, near-incompressible deformation, and the irreversible evolution of constituent damage. The matrix occupies the interval $\Omega_{m,0}^{1\mathrm{D}}=(0,L)$, with a length of $L=20$ mm and a cross-sectional area of $A_m=1$ mm$^2$. An independently discretized fiber occupies $\Gamma_{f,0}^{1\mathrm{D}}=(a,b)$, extending from $a=2$ mm to $b=18$ mm with a radius of $r_f=0.03$ mm. The matrix is fixed at its left boundary, while a cyclic axial displacement is prescribed at the opposite end. Both fiber ends remain free of essential boundary conditions, such that reinforcement loading occurs exclusively through the distributed matrix--fiber coupling. The matrix is discretized using $80$ linear elements with piecewise-constant pressure interpolation, while the reinforcement is represented by $63$ independently discretized line elements. No initial crack or prescribed damage seed is introduced. The geometry and loading configuration are illustrated in Fig.~\ref{fig:reduced-bar-schematic}.\\

\begin{figure}[!htbp]
  \centering
  \begin{subfigure}[t]{0.46\textwidth}
    \centering
    \includegraphics[width=\linewidth]{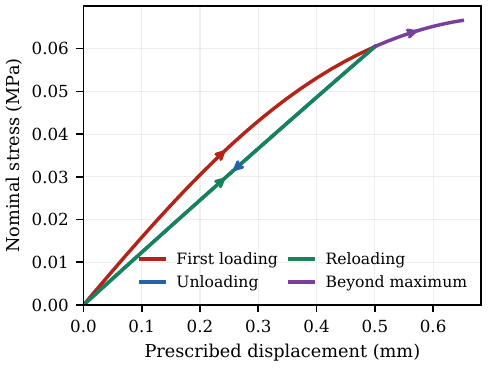}
    \caption{Nominal reaction stress.}
  \end{subfigure}\hfill
  \begin{subfigure}[t]{0.46\textwidth}
    \centering
    \includegraphics[width=\linewidth]{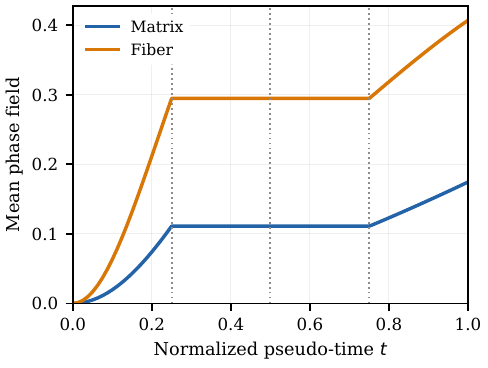}
    \caption{Mean constituent damage.}
  \end{subfigure}
  \caption{Cyclic response of the one-dimensional reinforced bar. The nominal reaction stress is shown during initial loading, unloading, reloading, and subsequent extension beyond the previous maximum displacement. The matrix and fiber damage histories represent spatially averaged phase fields over normalized pseudo-time. Vertical dotted lines indicate displacement reversals and the beginning of loading beyond the previous maximum. Damage remains unchanged during unloading and reloading below the previously attained maximum and resumes evolving upon further extension.}
  \label{fig:reduced-cyclic-response}
\end{figure}

The matrix is modeled as a nearly incompressible hyperelastic solid with shear modulus $\mu=1.00067$ MPa and volumetric coefficient $K=500$ MPa, corresponding to nominal elastic constants $E_m=3$ MPa and $\nu_m=0.499$. The reinforcement has a Young's modulus of $E_f=100$ MPa and Poisson's ratio $\nu_f=0.3$. Matrix fracture is characterized by $G_c=0.005$ N/mm, $\ell=0.5$ mm, and $k=10^{-6}$, while fiber fracture uses $G_{c,f}=10^{-4}$ N\,mm, $\ell_f=0.25$ mm, and $k_f=10^{-6}$. The constituents are connected through a translational penalty coupling with stiffness $k_{\mathrm{int}}=1000$ N/mm$^2$. The imposed displacement follows a loading--unloading--reloading sequence,
\begin{equation}
u_L:
0\longrightarrow0.50
\longrightarrow0
\longrightarrow0.50
\longrightarrow0.65~\mathrm{mm}.
\label{eq:reduced-cyclic-loading}
\end{equation}
This loading history permits the evolution of constituent damage to be examined during initial deformation, unloading, reloading, and subsequent extension beyond the previous maximum.\\
The reduced formulation retains five independent fields,
\begin{equation}
\mathbf{q}_{1\mathrm{D}}
=
\left[u_m,\,p_m,\,d_m,\,u_f,\,d_f\right]^T.
\label{eq:reduced-unknowns}
\end{equation}
Matrix and fiber displacement and damage fields are approximated using continuous linear interpolation, while the matrix pressure is represented by piecewise-constant interpolation. All fields are solved simultaneously using the monolithic Newton method, with the matrix and fiber irreversibility constraints enforced through the active-set procedure described in Sec.~\ref{sec:nonlinear-solution}.\\
To retain the three-dimensional constitutive behavior of the nearly incompressible matrix, transverse contraction is incorporated through a local constitutive reduction. A strictly one-dimensional incompressibility condition would prevent axial extension because the axial stretch would be constrained to unity. Instead, each matrix point is represented by a traction-free three-dimensional uniaxial deformation state,
\begin{equation}
\mathbf{F}_{1\mathrm{D}}
=
\operatorname{diag}(\lambda,\lambda_t,\lambda_t),
\qquad
\lambda=1+u_m'.
\label{eq:reduced-deformation-gradient}
\end{equation}
The transverse stretch is determined locally by enforcing the vanishing transverse nominal stresses and the mixed pressure equation,
\begin{equation}
\lambda_t^2
=
\frac{g(d_m)\mu}
{g(d_m)\mu-r(d_m)p_m\lambda},
\qquad
J=\lambda\lambda_t^2,
\qquad
0=-r(d_m)(J-1)-\frac{p_m}{K}.
\label{eq:reduced-condensation}
\end{equation}
This reduction allows the matrix to undergo axial extension with the corresponding transverse contraction while preserving the pressure-dependent constitutive response. The remaining constituent mechanics, fracture energies, and coupling terms follow the same formulation used in the multidimensional problems.\\
The resulting nominal reaction stress and evolution of the constituent phase fields are presented in Fig.~\ref{fig:reduced-cyclic-response}. During initial loading, both constituents accumulate damage as the imposed displacement increases. The fiber develops a larger mean phase field than the matrix, reflecting the different fracture resistance and deformation contributions of the two constituents. Damage accumulation progressively reduces the effective structural stiffness, producing a nonlinear force--displacement response. Upon unloading, the reaction decreases along a path with reduced stiffness relative to the initial loading response, consistent with the mechanical degradation accumulated during the preceding deformation.\\

The phase-field histories demonstrate the enforcement of constituent-wise irreversibility throughout the loading cycle. Although the mechanical deformation is reversed during unloading, neither damage field decreases, and both remain fixed during reloading below the previously attained maximum displacement. The subsequent reloading response therefore follows the degraded structural stiffness established during initial loading. Once the previous maximum displacement is exceeded, the matrix and fiber phase fields resume increasing, accompanied by additional changes in the reaction response. This behavior demonstrates that the monolithic active-set procedure preserves accumulated damage independently in both constituents while permitting reversible mechanical deformation about the damaged configuration.
The same calculation is used to examine the sensitivity of the matrix--fiber coupling to the translational penalty stiffness. Additional simulations are performed with $k_{\mathrm{int}}=500$ and $2000$ N/mm$^2$, corresponding to half and twice the baseline stiffness, while all remaining parameters are held fixed. Increasing the penalty stiffness systematically decreases the maximum matrix--fiber displacement gap, from $5.931\times10^{-5}$ mm to $1.496\times10^{-5}$ mm across the considered range. Despite this reduction, the terminal reaction remains approximately $0.06664$ N, with a relative variation below $6\times10^{-6}$, while the maximum fiber phase field varies by approximately $0.8\%$. 
These results demonstrate effective matrix–fiber load transfer with limited sensitivity of the global reaction and fiber damage to the selected penalty stiffness.
Together, these results provide a reduced-dimensional verification of several features of the proposed formulation. The independent matrix and fiber discretizations permit reinforcement loading through distributed coupling, while the mixed constitutive reduction accommodates axial extension under near-incompressible matrix behavior. The cyclic response demonstrates the irreversible evolution of both phase fields and the corresponding reduction in structural stiffness, with limited sensitivity to the selected coupling penalty. 

\subsection{Extension and fracture of an embedded helix}
\label{sec:helix-results}
\begin{figure}[!htbp]
  \centering
  \includegraphics[width=0.84\textwidth]{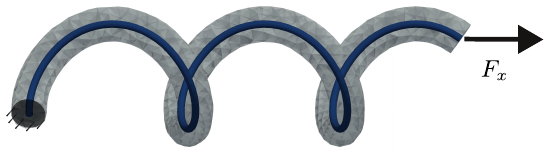}
  \caption{Undeformed helical reinforcement embedded within a conforming matrix tube. The left matrix end cap is clamped, while the opposite cap is subjected to a generalized axial force. The translucent matrix reveals the independently discretized beam and its nonmatching coupling to the surrounding solid.}
  \label{fig:helix-undeformed}
\end{figure}

\begin{figure}[htpb]
  \centering
  \includegraphics[width=\textwidth]{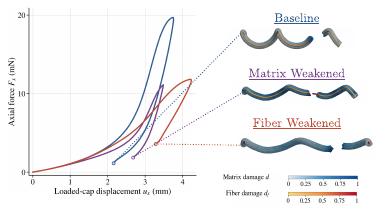}
  \caption{Axial force--displacement response and damage progression for the baseline, matrix-weakened, and fiber-weakened helices. Curves follow the accepted pseudo-arc-length continuation order. Open circles and matching dotted lines identify the first saved states satisfying the joint through-break criterion. The deformed configurations share a common viewing direction and fixed damage scales. Matrix and beam elements containing nodes above the damage threshold are omitted from the displayed configurations for visualization only; all elements remain active in the numerical calculation.}
  \label{fig:helix-progression}
\end{figure}

The extension and fracture of a spatial helical beam embedded within a conforming matrix tube is considered to examine coupled matrix--fiber damage under finite deformation and rotation. Unlike a straight reinforcement subjected to axial loading, the helical geometry produces a deformation response involving changes in curvature, cross-sectional rotation, and axial extension. The reference helix has a centerline radius of $0.35$ mm, extends $3.6$ mm along its axis, and completes $2.5$ turns. The surrounding matrix tube has a cross-sectional radius of $0.12$ mm, while the embedded beam has a radius of $0.025$ mm. The matrix is discretized using $2{,}850$ linear tetrahedral elements, and the reinforcement is represented by $120$ spatial finite-rotation beam elements. This configuration provides a three-dimensional example in which the matrix and reinforcement undergo coupled deformation while retaining independent discretizations.\\
The matrix is modeled as a nearly incompressible hyperelastic solid with shear and bulk moduli of $\mu=1$ MPa and $K=100$ MPa, respectively. The reinforcement has a Young's modulus of $E_f=100$ MPa and Poisson's ratio of $\nu_f=0.3$. The matrix and fiber phase-field regularization lengths are $\ell=0.12$ mm and $\ell_f=0.08$ mm, respectively. To investigate the influence of constituent fracture resistance, three configurations are considered: a baseline case, a matrix-weakened case, and a fiber-weakened case. Their fracture parameters are
\begin{equation}
(G_c,G_{c,f})=
\begin{cases}
(0.05~\mathrm{N/mm},\,5\times10^{-4}~\mathrm{N\,mm}), & \text{baseline},\\
(0.005~\mathrm{N/mm},\,5\times10^{-4}~\mathrm{N\,mm}), & \text{matrix-weakened},\\
(0.05~\mathrm{N/mm},\,1.1\times10^{-5}~\mathrm{N\,mm}), & \text{fiber-weakened}.
\end{cases}
\label{eq:helix-fracture-parameters}
\end{equation}
Only the fracture resistance of the designated constituent is modified between cases, while the geometry, elastic properties, coupling parameters, and discretization remain unchanged. This allows differences in the resulting response to be associated with the prescribed changes in constituent fracture resistance.\\
One matrix end cap is clamped, while the opposite cap is subjected to a generalized axial force,
\begin{equation}
F_x=\Lambda F_{\mathrm{ref}},
\end{equation}
where $\Lambda$ is the load factor and $F_{\mathrm{ref}}$ is the reference force used in the continuation procedure. The loaded cap has a common unknown axial translation, while its transverse displacements remain free. Matrix damage is prescribed to vanish on both end caps, and no essential boundary conditions are imposed directly on the beam ends. The undeformed configuration and loading arrangement are illustrated in Fig.~\ref{fig:helix-undeformed}.\\
Pseudo-arc-length continuation is employed to follow the equilibrium response through structural softening and snap-back \cite{crisfield1981fast}. The load factor is solved simultaneously with the mechanical, pressure, and damage fields, allowing the equilibrium path to pass through force maxima and onto descending or returning branches. The reported force is the discrete reaction conjugate to the axial translation of the loaded cap. Although the arc-length increment controls progression between computed equilibrium states, the corresponding force and axial displacement provide the physical measures of structural resistance and deformation. Consequently, progression along the continuation path does not necessarily imply increasing axial extension, particularly after localized fracture produces a loss of structural stiffness.\\
Figure~\ref{fig:helix-progression} compares the axial force--displacement responses and selected damage configurations for the three cases. The initial responses are similar because the geometry and elastic properties are identical. As damage develops, the equilibrium paths diverge, reflecting the influence of constituent fracture resistance on the evolving structural response. The baseline configuration reaches the highest maximum force, while reductions in either matrix or fiber fracture resistance decrease the maximum structural resistance and modify the subsequent softening behavior. All three configurations exhibit pronounced nonlinear responses beyond their respective force maxima, including descending and returning equilibrium branches. These features indicate that the degradation of the individual constituents affects not only the maximum resistance but also the progression of fracture throughout the coupled structure.\\
To distinguish the initiation of strongly localized damage from the subsequent loss of structural connectivity, two fracture indicators are considered. The first local damage event is identified when either constituent reaches the prescribed threshold,
\begin{equation}
\max_{\Omega_0}d>0.95
\qquad\text{or}\qquad
\max_{\Gamma_{f,0}}d_f>0.95.
\label{eq:helix-local-damage-threshold}
\end{equation}
 In the baseline and fiber-weakened configurations, the first threshold crossing occurs in the reinforcement, whereas the matrix-weakened configuration first reaches the threshold in the surrounding matrix. Thus, reducing the fracture resistance of an individual constituent can change which part of the coupled structure develops strongly localized damage first. Importantly, the initial threshold crossing does not imply complete structural separation, because the other constituent can continue transmitting load between the specimen ends.\\
The distinction between the initial damage event and the subsequent structural response highlights the interaction between the two fracture fields. When fiber damage develops first, the surrounding matrix continues to provide resistance and transfers loading around the degraded reinforcement. Conversely, when matrix damage develops first, the embedded beam continues to bridge the damaged region and contributes to the remaining structural resistance. The deformation of the helical reinforcement is particularly important in this process because the beam can accommodate the evolving matrix deformation through combinations of axial stretching, bending, and rotation. The resulting response is therefore governed by the coupled deformation and degradation of both constituents rather than by the fracture resistance of either constituent independently.\\
To characterize the subsequent progression toward separation, a joint through-break is identified as the first saved equilibrium state in which neither constituent retains a continuous subthreshold path between the two axial ends. Specifically, the matrix region satisfying $d<0.95$ and the fiber region satisfying $d_f<0.95$ must both lose end-to-end connectivity. This criterion is evaluated using the native piecewise-linear phase fields and permits the development of nonplanar or oblique fracture paths. The corresponding states are indicated by open circles in Fig.~\ref{fig:helix-progression}, together with the associated deformed configurations.
In all three cases, the joint through-break occurs after the maximum structural resistance has been reached, demonstrating that substantial additional damage develops during the post-peak response. The marked states are characterized by considerably reduced reaction forces, although their corresponding displacements differ because the equilibrium paths undergo distinct softening and snap-back excursions. In particular, the axial extension at the joint through-break state may be smaller than at an earlier local damage event. This does not indicate a reversal of fracture: damage remains irreversible throughout the continuation procedure, while the equilibrium displacement is permitted to decrease as the structure follows a returning branch. The pronounced differences between the three responses further demonstrate that changing the fracture resistance of one constituent influences the subsequent equilibrium path of the coupled system.\\
The displayed damage configurations provide a complementary view of this progression. The loss of strongly damaged material from the visualization exposes the developing separation regions and illustrates how the matrix and reinforcement deform as fracture advances. These regions should not be interpreted as explicitly removed material: the underlying finite elements, residual stiffness contributions, and matrix--fiber coupling remain present in the mechanical problem.\\
Together, these results demonstrate that the proposed monolithic formulation can capture coupled matrix and fiber fracture in a geometrically complex structure undergoing finite rotations and substantial deformation. The comparison illustrates how variations in constituent fracture resistance alter the sequence of localized damage, the maximum structural resistance, and the subsequent post-peak equilibrium response. Furthermore, the ability to follow the system beyond its maximum force enables the evolution from initial localized degradation to joint loss of subthreshold connectivity to be examined within a single continuous solution framework.\\

\subsection{Deformation and damage of a periodic fiber network}
\label{sec:periodic-rve-results}
\begin{figure}[!htbp]
  \centering
  \includegraphics[width=0.5\textwidth]{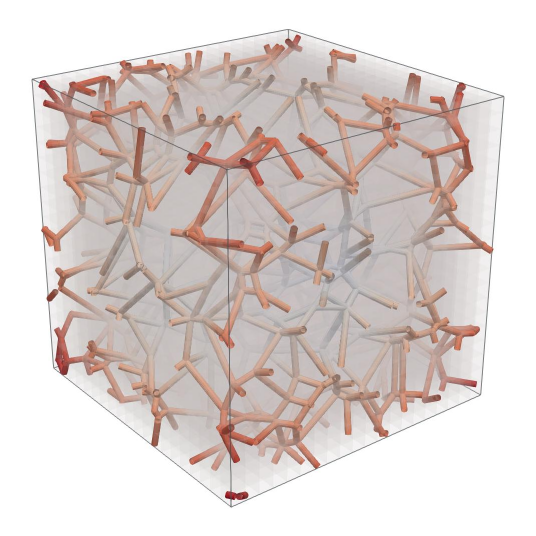}
  \caption{Undeformed periodic unit cell containing an irregular three-dimensional fiber network generated using a periodic Voronoi tessellation. The reinforcement is represented by finite-rotation beams embedded within the translucent matrix. Fiber colors indicate distance from the cell center to distinguish the network geometry and do not represent a mechanical field.}
  \label{fig:periodic-rve-initial-mesh}
\end{figure}

\begin{figure}[htpb]
  \centering
  \begin{subfigure}[t]{0.48\textwidth}
    \centering
    \includegraphics[width=0.95\textwidth]{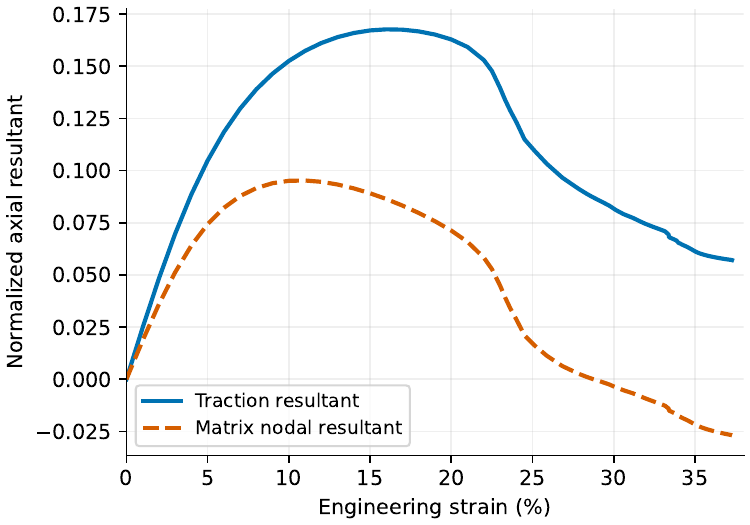}
    \caption{Axial-resultant diagnostics.}
  \end{subfigure}\hfill
  \begin{subfigure}[t]{0.48\textwidth}
    \centering
    \includegraphics[width=0.95\textwidth]{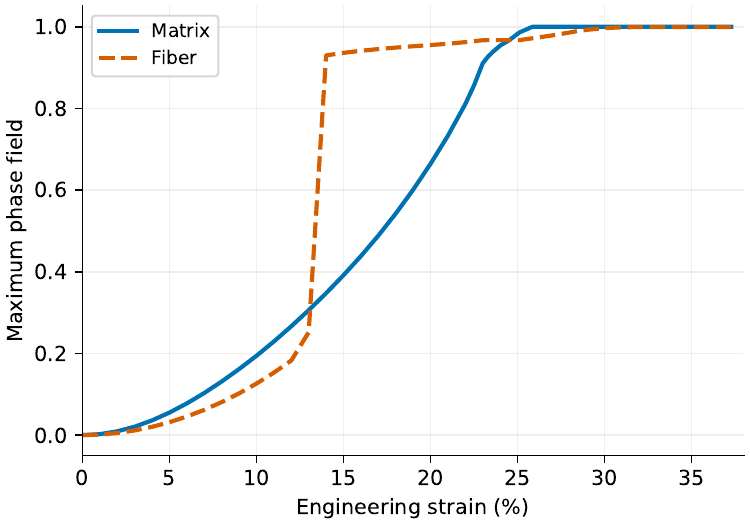}
    \caption{Maximum constituent damage.}
  \end{subfigure}
  \begin{subfigure}[t]{0.48\textwidth}
    \centering
    \includegraphics[width=0.95\textwidth]{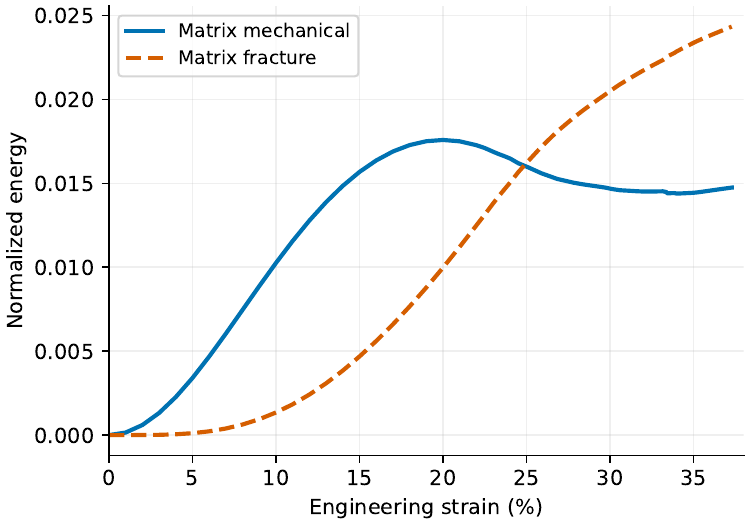}
    \caption{Matrix energy contributions.}
  \end{subfigure}\hfill
  \begin{subfigure}[t]{0.48\textwidth}
    \centering
    \includegraphics[width=0.95\textwidth]{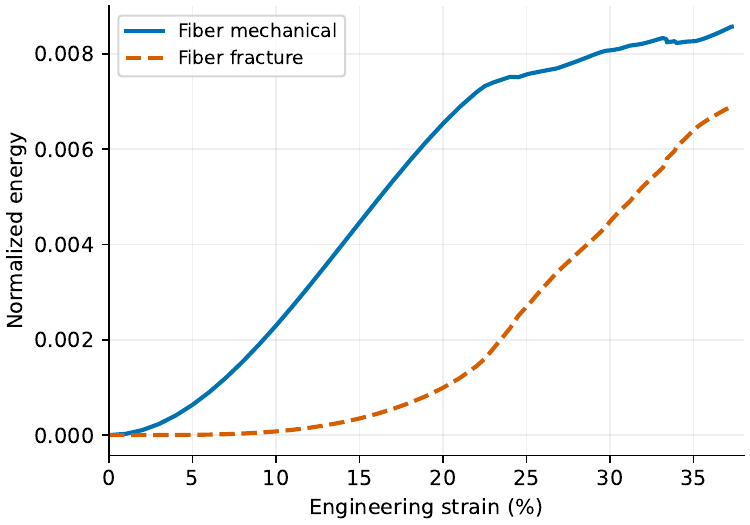}
    \caption{Fiber energy contributions.}
  \end{subfigure}
  \captionsetup{skip=4pt}
  \caption{Periodic-cell response: (a) axial resultants, (b) maximum constituent damage, and (c,d) matrix and fiber energies. Curves show record-high extension states, omitting returning branches. Coupling energies are excluded.}
  \label{fig:periodic-rve-damage}
\end{figure}

\begin{figure}[!htbp]
  \centering
  \begin{subfigure}[t]{0.48\textwidth}
    \centering
    \includegraphics[width=\textwidth]{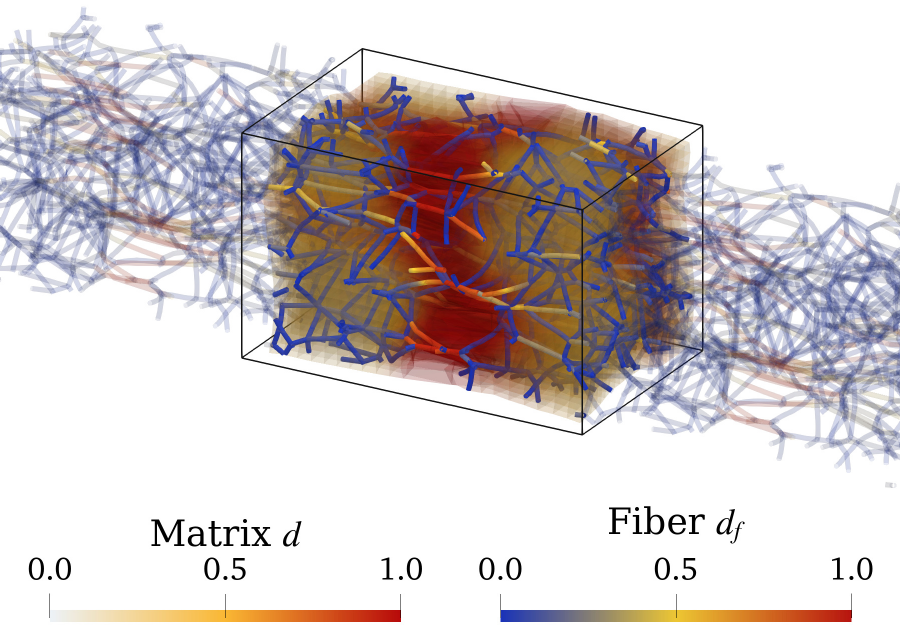}
    \caption{Matrix and fiber phase fields.}
  \end{subfigure}\hfill
  \begin{subfigure}[t]{0.48\textwidth}
    \centering
    \includegraphics[width=\textwidth]{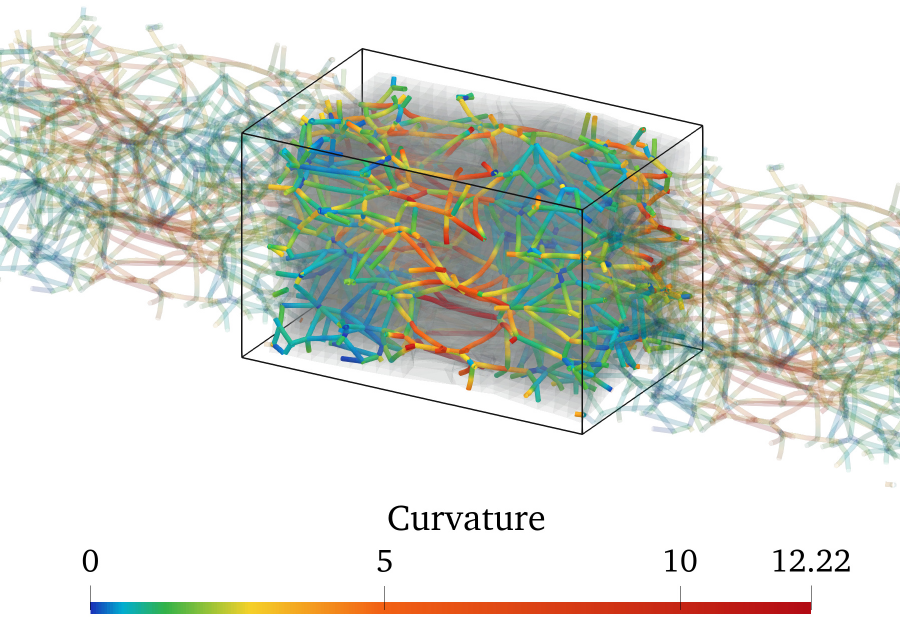}
    \caption{Fiber bending curvature.}
  \end{subfigure}
  \caption{Spatial distribution of matrix damage, fiber damage, and reinforcement bending near the final reported axial extension. The constituent damage fields use the common range $0\leq d,d_f\leq1$, while the normalized bending-curvature magnitude is shown using a fixed visualization scale with its upper limit corresponding to the 99th percentile. The matrix is rendered semitransparently, and the fibers are displayed at their modeled radii. Neighboring periodic copies illustrate the continuity of the reinforcement across the unit-cell boundaries. Deformation is shown without amplification.}
  \label{fig:periodic-rve-field}
\end{figure}

The deformation and damage of an irregular three-dimensional fiber network embedded within a periodic matrix cell are considered to examine the coupled response of a spatially distributed reinforcement system under finite deformation. The network is constructed using a periodic Voronoi tessellation, with matching branches across opposing cell faces to ensure continuity of the reinforcement under periodic boundary conditions \cite{Thompson2002Periodic,Yan2013ClippedVoronoi}. The resulting network contains $\sim2{,}000$ native branches with a mean coordination number of 2.973, discretized using $\sim4{,}000$ spatial finite-rotation beam elements and corresponding to a nominal superposed reinforcement volume fraction of $4.67\%$. The surrounding matrix is discretized using $20\times20\times20$ hexahedral elements. The initial configuration is illustrated in Fig.~\ref{fig:periodic-rve-initial-mesh}.
The cell dimensions and constituent properties are normalized using the reference cell length $L_0$ and matrix shear modulus $\mu_0$. The periodic cell is represented by a unit cube, with $L_0=1$ in the nondimensional configuration. The corresponding force and energy scales are $\mu_0L_0^2$ and $\mu_0L_0^3$, respectively. Because matrix and fiber fracture are regularized over domains of different dimensions, their fracture-energy parameters require distinct normalizations. Using the reference cell length $L_0$ and matrix shear modulus $\mu_0$, the dimensionless fracture energies are defined as
\begin{equation}
    \widehat{G}_c=\frac{G_c}{\mu_0L_0},
    \qquad
    \widehat{G}_{c,f}=\frac{G_{c,f}}{\mu_0L_0^3}.
    \label{eq:normalized-fracture-energies}
\end{equation}
The dimensionless material and regularization parameters employed in the periodic-cell calculation are
\begin{equation}
\begin{gathered}
\frac{K}{\mu_0}=1000,\quad
\frac{E_f}{\mu_0}=100,\quad
\nu_f=0.3, \quad
\frac{r_f}{L_0}=
\widehat{G}_c=0.01,\quad
\widehat{G}_{c,f}=10^{-4},\quad
\frac{\ell}{L_0}=0.1,\quad
\frac{\ell_f}{L_0}=0.05.
\end{gathered}
\label{eq:periodic-rve-parameters}
\end{equation}
The reinforcement is modeled using spatial beams that support axial extension, transverse shear, bending, and torsion. Network junctions are represented as ideal hinges that transfer forces while permitting independent branch rotations. Although moments are not transmitted directly through these junctions, individual branches can develop bending and torsion through their interaction with the surrounding matrix. The nonmatching matrix--fiber coupling transfers the corresponding deformation and internal forces between the constituents.\\
The periodic cell is subjected to macroscopic isochoric axial extension, prescribed through the deformation gradient
\begin{equation}
\overline{\mathbf{F}}
=
\operatorname{diag}
\left(
\lambda,\lambda^{-1/2},\lambda^{-1/2}
\right),
\qquad
\lambda=1+\frac{1}{2}\Lambda,
\label{eq:periodic-rve-loading}
\end{equation}
where $\Lambda$ denotes the macroscopic loading parameter and the corresponding engineering axial extension is $\lambda-1=\Lambda/2$. Following standard first-order periodic homogenization \cite{kouznetsova2001micro,saeb2016homogenization}, affine-periodic displacements are imposed on the matrix and fiber centerlines, while the pressure, matrix and fiber phase fields, and beam rotational fluctuations satisfy periodic conditions across opposing faces. The prescribed macroscopic deformation is volume-preserving, while the periodic fluctuations permit the matrix and individual fiber branches to undergo nonuniform local deformation. Periodic constraints connect the clipped fiber segments across opposing cell boundaries.\\
An initial displacement-controlled stage is followed by pseudo-arc-length continuation to trace the coupled equilibrium response through progressive damage and structural softening. The accepted history includes returning equilibrium branches and extends to approximately $37\%$ macroscopic axial extension. \\
Figure~\ref{fig:periodic-rve-damage} presents the evolution of the axial-resultant diagnostics, maximum constituent damage, and matrix and fiber energy contributions. Both axial resultants initially increase, reach distinct maxima, and subsequently decrease with continued deformation. The matrix nodal resultant reaches its maximum earlier and eventually becomes negative, whereas the combined traction-based resultant remains positive throughout the reported extension range. These quantities are evaluated differently: the matrix nodal resultant is obtained from the assembled residual on the loaded matrix face, while the traction-based resultant combines the matrix-face constitutive traction with the periodic fiber boundary contribution. The decreasing traction-based resultant is consistent with progressive softening of the coupled system as constituent damage accumulates.\\
The evolution of the constituent phase fields reveals distinct stages of damage development within the matrix and reinforcement. At small macroscopic extensions, the maximum matrix phase field exceeds that of the fibers. The fiber maximum then increases sharply and becomes larger over an intermediate range of deformation. As extension continues, matrix damage increases more rapidly and again exceeds the fiber maximum. Both constituent maxima approach unity near the end of the reported extension range, but the onset and progression of severe local degradation differ between the constituents. These trends demonstrate that the relative evolution of constituent damage is nonuniform throughout loading. However, the reported maxima characterize local degradation and do not independently establish the formation of a continuous fracture surface or the complete loss of structural load-bearing capacity.\\
The matrix energy contributions in Fig.~\ref{fig:periodic-rve-damage}(c) provide further insight into the evolving mechanical response. Initially, the matrix mechanical energy increases as deformation accumulates, reflecting the storage of recoverable energy within the surrounding solid. With continued extension, the mechanical energy reaches a maximum and subsequently declines overall, with a slight recovery near the final reported states, while the matrix fracture energy increases progressively and eventually exceeds the mechanical contribution. This contrasting evolution is consistent with the accumulation of matrix damage and the accompanying reduction in its capacity to store mechanical energy.\\
The fiber energy contributions exhibit a different evolution, as shown in Fig.~\ref{fig:periodic-rve-damage}(d). Unlike the matrix, the fiber mechanical energy increases overall throughout the reported extension range, with minor local fluctuations, even as the axial resultants decrease. The fiber fracture energy also increases progressively, indicating continued degradation of the reinforcement. These contrasting energetic responses demonstrate that the matrix and reinforcement accommodate the evolving deformation through distinct mechanisms. In particular, the continued accumulation of fiber mechanical energy indicates that reinforcement deformation remains significant during the progressive softening of the surrounding matrix.\\
The spatial distributions of matrix damage, fiber damage, and reinforcement bending are presented in Fig.~\ref{fig:periodic-rve-field}. The deformed network accommodates macroscopic extension through a combination of axial stretching, finite rotations, and changes in branch curvature. Although the prescribed macroscopic deformation is homogeneous, the irregular network architecture produces heterogeneous local deformation, with larger bending curvatures concentrated along selected branches. These nonuniform deformation patterns are accommodated by the spatial beam formulation, which retains the axial, shear, bending, and torsional contributions of individual reinforcement members under finite rotations.\\
The damage distributions further demonstrate the interaction between the irregular reinforcement and the surrounding matrix. Matrix degradation develops over a relatively broad region of the cell, whereas fiber damage is concentrated along selected network members. These differences arise from the distinct deformation measures and fracture-driving contributions of the two constituents. In particular, bending contributes to the axial section strain defined in Eq.~\eqref{eq:fiber-section-strains}, which enters the tensile fracture-driving energy in Eq.~\eqref{eq:fiber-energy-split}, allowing changes in bending curvature to contribute directly to fiber degradation. Torsion also contributes to the mechanical response and redistribution of internal forces, although it does not directly enter the adopted tensile fracture-driving energy. Consequently, individual branches may experience substantial rotational deformation without developing equivalent levels of fiber damage. The resulting response reflects the combined influence of network geometry, beam kinematics, and the mechanical constraints imposed by the surrounding matrix.\\
The periodic deformation fields further demonstrate the continuity of the reinforcement across the cell boundaries. Although the network contains clipped branches at opposing faces, the periodic constraints permit the corresponding fibers to deform consistently across neighboring cells. The prescribed macroscopic deformation remains isochoric, while the local matrix and reinforcement fields accommodate heterogeneous deformation and progressive degradation. Together, these results demonstrate the ability of the coupled formulation to resolve distinct matrix and fiber damage mechanisms within an irregular three-dimensional periodic network. The differing evolution of the constituent phase fields and energy contributions further highlights the importance of resolving their mechanical and fracture responses independently.

\subsection{Mode-I fracture of a fiber-reinforced solid}
\label{sec:type-i-4000-results}
\begin{figure}[!htbp]
  \centering
  \begin{minipage}[c]{0.44\textwidth}
    \centering
    \includegraphics[height=1.29\linewidth]{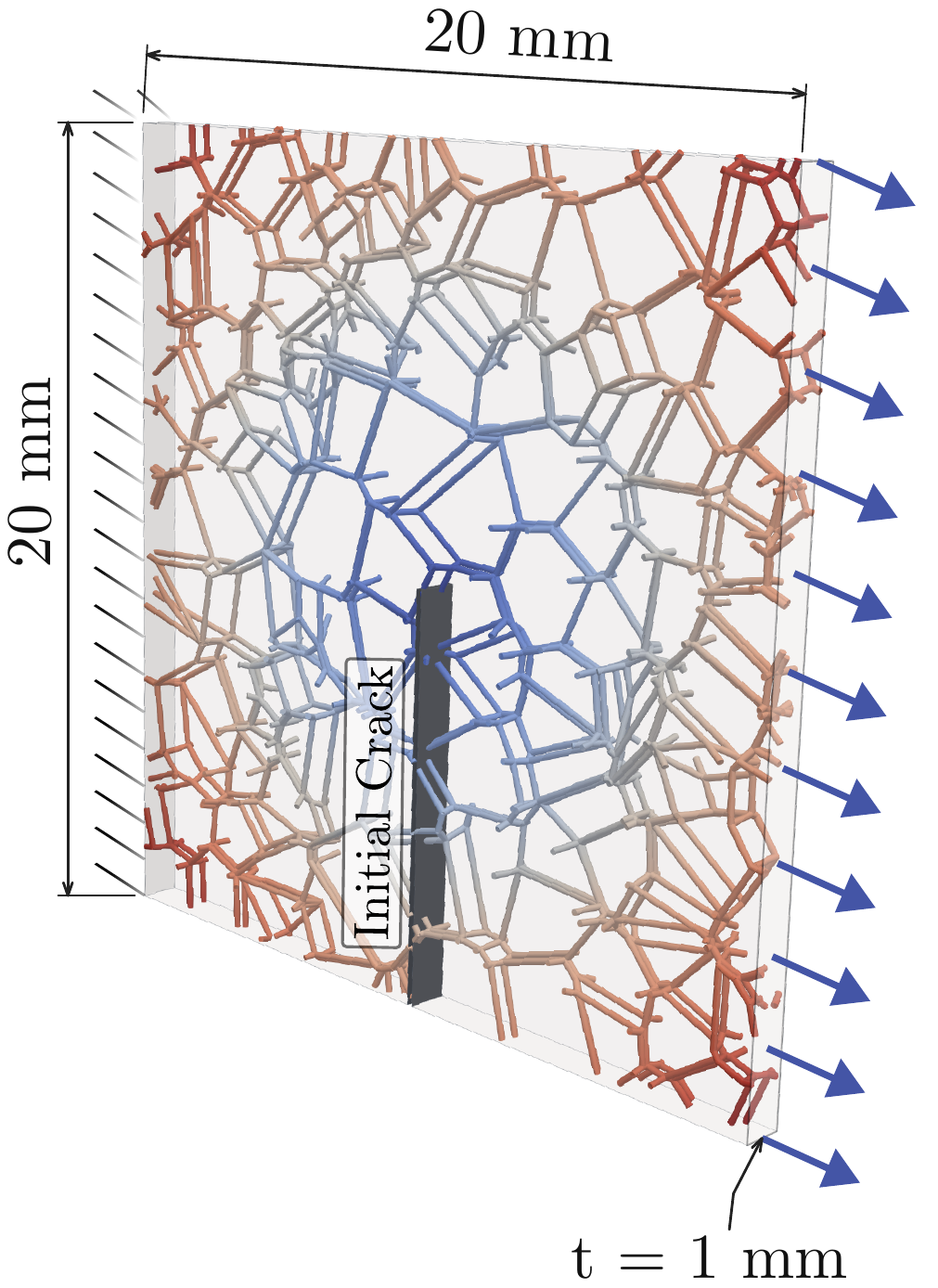}
  \end{minipage}\hfill
  \begin{minipage}[c]{0.54\textwidth}
    \centering
    \includegraphics[height=1.051111\linewidth]{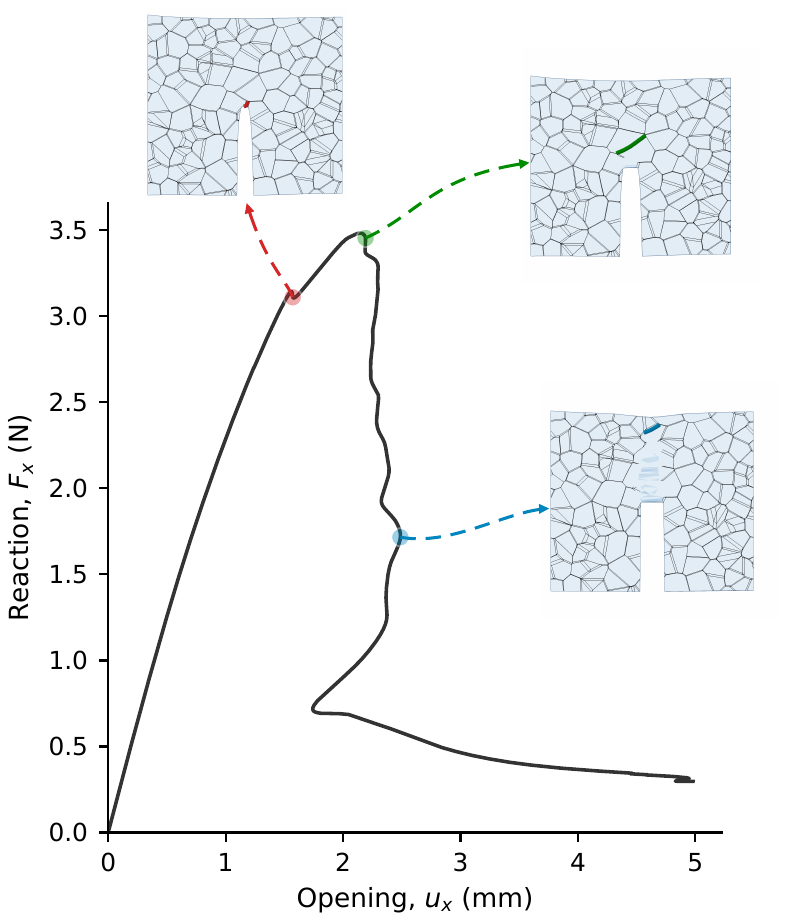}
  \end{minipage}
  \begin{subcaptiongroup}
    \phantomcaption\label{fig:type-i-event-curve}
    \phantomcaption\label{fig:type-i-4000-matrix-progression}
    \phantomcaption\label{fig:type-i-4000-fiber-progression}
  \end{subcaptiongroup}
  \caption{Undeformed lower-volume-fraction network (left) and corresponding reaction--opening response with selected fiber-damage events (right). Fiber colors in the reference configuration indicate distance from the specimen center. Colored markers identify equilibrium states at which the highlighted fiber segments first reach the critical damage threshold, with dashed arrows indicating their locations within the network.}
  \label{fig:type-i-4000-response}
\end{figure}

Mode-I fracture of an edge-notched solid reinforced by an irregular three-dimensional fiber network is considered to examine the interaction between matrix crack propagation, progressive fiber degradation, and redistribution of load through the surrounding reinforcement. The specimen is a square plate with in-plane dimensions of $20\times20$ mm and a thickness of $1$ mm. A through-thickness edge notch, located at the midpoint of one edge, extends $10$ mm into the specimen and has a width of $0.15$ mm. The left face is fixed, while an axial opening displacement is prescribed on the opposite face. Two embedded fiber networks with nominal reinforcement volume fractions of $4.93\%$ and $43.75\%$ are considered, comprising $\sim 1{,}600$ and $\sim 40{,}000$ native branches discretized by $\sim 8{,}000$ and $\sim85{,}000$ beam elements, respectively. Despite the substantial difference in network size, their mean branch coordination numbers remain comparable at $2.452$ and $2.377$, respectively. Both calculations employ the same matrix discretization of $\sim 14{,}000$ tetrahedral elements, with beam-element lengths limited to $0.15$ mm.\\
Following an initial displacement-controlled loading stage, pseudo-arc-length continuation is employed to trace the coupled equilibrium response through structural softening and snap-back. The applied opening is parameterized by
\begin{equation}
u_x=\lambda u_{\mathrm{ref}},
\end{equation}
where the load factor $\lambda$ is solved simultaneously with the mechanical and damage fields. Unlike conventional monotonic displacement control, pseudo-arc-length continuation permits the equilibrium opening to increase or decrease as the structure progresses along its equilibrium path. The arc-length increment determines the numerical progression between equilibrium configurations, while the corresponding opening displacement and reaction provide physical measures of deformation and structural resistance. A weighted continuation constraint incorporates the mechanical variables and both phase fields, with each increment obtained using a tangent predictor and monolithic Newton corrector. Damage irreversibility is maintained throughout the continuation procedure, allowing progressive fracture to be followed through regions of negative incremental stiffness and returning equilibrium branches.\\
We first consider the lower-volume-fraction network to examine how localized fiber degradation influences the global structural response. The undeformed reinforcement geometry and corresponding reaction--opening response are presented in Fig.~\ref{fig:type-i-4000-response}, together with selected fiber-damage events along the equilibrium path. During initial loading, the reaction increases despite the onset of localized fiber degradation, indicating that the surrounding matrix and remaining reinforcement continue to accommodate increasing loads. The first highlighted damage event occurs along the ascending equilibrium branch, followed by additional events near the maximum reaction and during subsequent softening. Beyond the peak, the response becomes strongly nonmonotonic, exhibiting pronounced reaction fluctuations and successive snap-back excursions. These features are consistent with localized reductions in reinforcement stiffness and the accompanying redistribution of internal forces among the remaining load-bearing branches. The sequence of highlighted events illustrates how fiber degradation develops progressively rather than through simultaneous failure of the network.\\

\begin{figure}[!htbp]
  \centering
  \includegraphics[width=\linewidth]{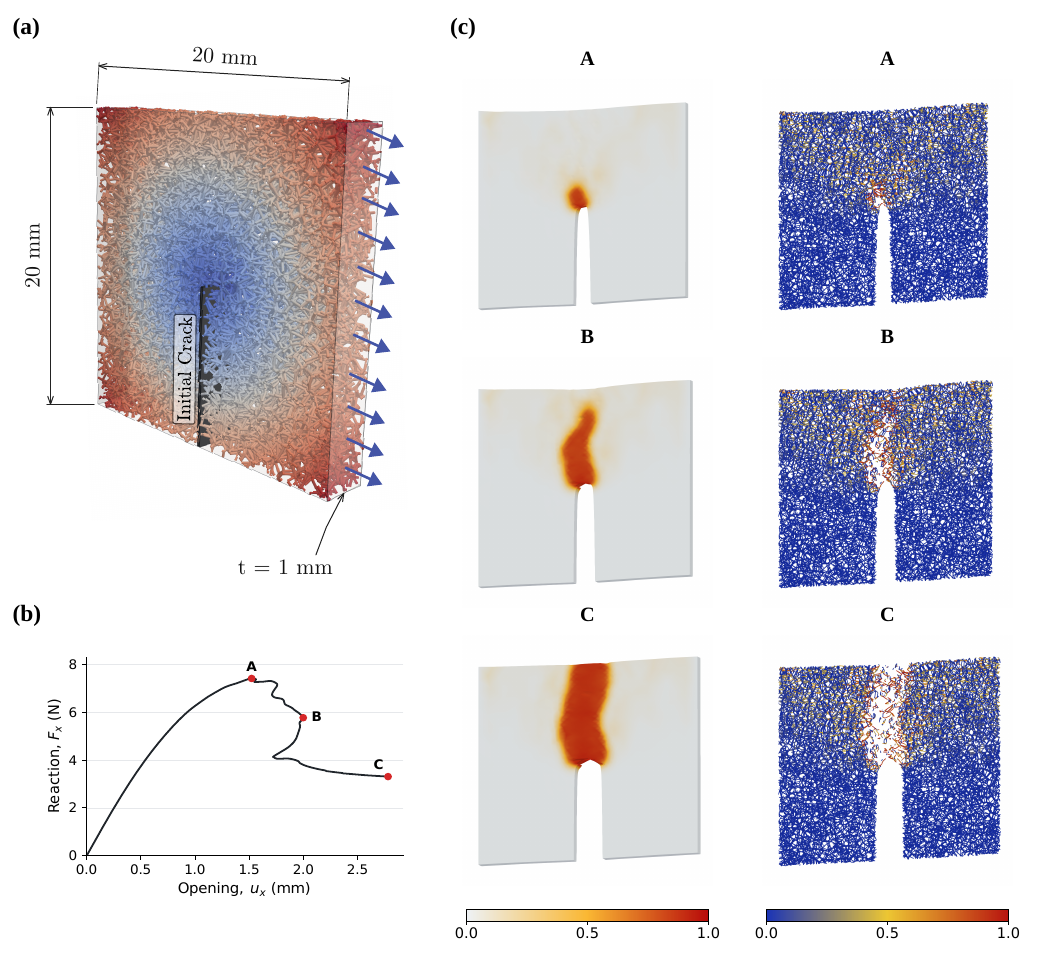}
  \caption{Mode-I fracture of the higher-volume-fraction fiber-reinforced solid. (a) Edge-notched specimen geometry and loading configuration, with the left face fixed and an axial opening displacement prescribed on the opposite face. (b) Reaction--opening equilibrium path with selected states A--C. (c, left column) Matrix damage progression and (c, right column) fiber damage progression at states A--C from top to bottom. Matrix and fiber damage are displayed on fixed zero-to-one scales. Matrix and fiber elements with nodal damage exceeding $0.95$ are omitted for visualization only. Three retraced continuation excursions are omitted from the displayed reaction curve for clarity.}
  \label{fig:type-i-dense-damage}
  \label{fig:type-i-loading}
\end{figure}

The pronounced reaction fluctuations highlight the influence of individual reinforcement failures on the response of the lower-volume-fraction network. As a load-bearing segment undergoes substantial degradation, its ability to transmit force diminishes, requiring neighboring branches and the surrounding matrix to accommodate the redistributed loading. This redistribution can preserve residual structural resistance while increasing the deformation and fracture-driving forces experienced by other network members. Consequently, additional localized degradation may develop as the equilibrium path progresses, producing successive changes in structural stiffness. Importantly, the underlying fiber graph may remain geometrically connected even as strongly damaged segments cease to contribute appreciably to its mechanical resistance. The effective load-bearing connectivity therefore evolves independently of the geometric topology, allowing localized failure within a relatively short segment to compromise the reinforcing contribution of a larger connected fiber path. The associated reaction fluctuations and snap-back excursions demonstrate the complex equilibrium behavior that can accompany progressive reinforcement degradation, although the contributions of individual fiber failures cannot be isolated from the global reaction curve alone.\\
We next consider the higher-volume-fraction network to examine how these interacting fracture mechanisms develop within a substantially denser reinforcement architecture. Figure~\ref{fig:type-i-dense-damage} presents the specimen geometry, reaction--opening response, and evolution of matrix and fiber damage at selected equilibrium states. The larger network provides many interconnected reinforcement paths through the surrounding matrix, allowing forces to be redistributed as individual members degrade. During initial loading, the reaction increases as the embedded fibers contribute to the structural resistance and transfer load across the developing fracture region. Following the maximum reaction, the specimen undergoes substantial structural softening while retaining residual resistance through the remaining matrix and less-damaged reinforcement. Compared with the lower-volume-fraction network, individual damage events are less prominent in the global reaction, consistent with a more distributed load-sharing response.\\

The spatial evolution of the constituent phase fields reveals how matrix cracking and fiber degradation interact as fracture progresses. Matrix damage initially concentrates near the notch tip and subsequently extends into the remaining ligament, producing an increasingly developed fracture region. The corresponding fiber-damage field evolves along reinforcement branches intersecting and surrounding this region, reflecting the progressive deterioration of the load-bearing network. Between the selected equilibrium states, the matrix fracture band advances while fiber degradation develops over a broader portion of the surrounding reinforcement. These differences arise from the distinct fracture-driving mechanisms of the two constituents and the ability of the embedded network to transmit forces and moments through interconnected branches. The resulting damage distributions demonstrate that matrix crack propagation and fiber failure need not remain spatially coincident, even though their evolution is mechanically coupled.
An important consequence of this interaction is that fiber degradation is not necessarily confined to the immediate vicinity of the localized matrix fracture region. While matrix damage develops predominantly along the advancing fracture band, the embedded reinforcement transmits forces and moments through connected branches extending beyond it. Degradation of an individual fiber segment therefore modifies the internal force distribution throughout the surrounding network, potentially increasing deformation and fracture-driving forces in spatially separated members. This produces a mechanically nonlocal response in which fiber damage depends not only on the local matrix deformation but also on the connectivity and load-bearing contributions of neighboring branches. Unlike the nonlocal regularization introduced by the phase-field length scales, this redistribution arises directly from the mechanical interactions transmitted through the connected reinforcement. As damage accumulates, the network progressively reorganizes its load-bearing paths, allowing less-damaged branches to continue supporting the structure even after substantial localized fracture.
Comparison of the two reinforcement networks reveals distinct manifestations of progressive matrix--fiber fracture. In the lower-volume-fraction network, degradation of individual load-bearing branches has a pronounced influence on the structural response, producing substantial reaction fluctuations and returning equilibrium branches. In the higher-volume-fraction network, matrix cracking and fiber degradation remain strongly coupled, but the contributions of individual reinforcement failures are less apparent in the global reaction. The substantially greater number of available load-bearing paths is consistent with forces released by degrading segments being redistributed among more neighboring branches, reducing the prominence of individual failure events. Although the network configurations differ in both reinforcement content and architecture, their contrasting responses illustrate how the collective behavior of an interconnected network can differ substantially from the failure of its individual members. Together, these examples demonstrate the ability of the proposed formulation to capture progressive matrix--fiber fracture, mechanically nonlocal load redistribution, evolving load-bearing connectivity, and the nonlinear structural response associated with localized reinforcement failure.\\

\FloatBarrier
\section{Conclusions}
\label{sec:conclusions}
In this work, we developed a monolithic mixed-dimensional finite-element
framework for modeling fracture in nearly incompressible hyperelastic
solids containing embedded fiber networks. The formulation combines
a stabilized displacement--pressure description of the surrounding
matrix with finite-rotation beam elements and independent phase fields
for matrix and fiber damage. By coupling the constituents through
translational and rotational penalty terms, forces and moments can
be transferred across nonmatching discretizations without requiring
a conforming volumetric mesh of individual fibers. Moreover, solving
the mechanical and fracture equations simultaneously allows damage
in either constituent to influence the deformation and loading of
the other. We first examined this coupled response through cyclic
loading of a one-dimensional reinforced bar, demonstrating the
irreversible evolution of matrix and fiber damage. 
Building on this initial example, we investigated how interacting
matrix and fiber damage influences the structural response of
materials containing discrete networks. We first considered an
embedded helix, where variations in constituent fracture resistance
altered the sequence of damage development and the subsequent
post-peak response. Extending the analysis to an irregular periodic
network then allowed us to examine how finite rotations, bending,
and network architecture contribute to deformation and damage
throughout a connected three-dimensional structure. Finally,
we applied the framework to Mode-I fracture of a densely reinforced
solid, where the interaction between matrix cracking and fiber
degradation became particularly apparent. While matrix damage
localized near the crack tip, fiber damage developed across a
broader region of the surrounding network. Furthermore, as
damage progressed, fiber degradation was accompanied by changes
in the global reaction.
Together, these examples demonstrate how resolving individual
network members allows us to distinguish the failure processes
of the two constituents and investigate their influence on
the overall structural response. \\
While these examples demonstrate the capabilities of the proposed
formulation, several extensions would broaden its applicability.
The current framework assumes a bonded matrix--fiber interface,
represents fiber rupture through whole-section degradation,
and employs idealized network junctions. Future developments
could therefore incorporate interfacial debonding, more detailed
junction failure mechanisms, and experimentally informed
constituent models. Combined with further numerical investigations
and experimental validation, these extensions would provide
a basis for studying how network architecture, constituent
properties, and their interacting failure mechanisms influence
the progression of fracture and the loss of structural integrity
in complex soft materials.

\section*{Acknowledgements}
This material is based upon work partially supported by the U.S. National Science Foundation under award No. 2452029 (JNF) and award Nos. 2127925, 2235856, 2046148, and 2438943 (MKR). This work was also supported by the Office of Naval Research under awards N00014-25-1-2216, N00014-25-1-2237, and N00014-23-1-2575 (MKR)). Any opinions, findings, and conclusions or recommendations expressed in this material are those of the authors and do not necessarily reflect the views of the National Science Foundation or the Office of Naval Research.
\\
The authors acknowledge the Texas Advanced Computing Center (TACC) at The University of Texas at Austin for providing computational resources that have contributed to the research results reported within this paper.

\begingroup
\setlength{\itemsep}{0pt}
\setlength{\parsep}{0pt}
\bibliographystyle{unsrt}
\bibliography{references}
\endgroup

\appendix

\setcounter{table}{0}
\setcounter{figure}{0}
\renewcommand{\thetable}{\thesection.\arabic{table}}
\renewcommand{\theHtable}{\theHsection.\arabic{table}}
\renewcommand{\thefigure}{\thesection.\arabic{figure}}
\renewcommand{\theHfigure}{\theHsection.\arabic{figure}}

\end{document}